\documentclass[twocolumn]{aastex631}

\usepackage{fontawesome}
\usepackage{amsmath}	% Advanced maths commands
\usepackage{amssymb}	% Extra maths symbols
\usepackage{bm}

\usepackage{xcolor}
\usepackage{hyperref}
\usepackage[varg]{txfonts}
\usepackage{comment}
\usepackage{upgreek}
\let\tablenum\relax
\usepackage{siunitx}

\newcommand{\cs}{\ensuremath{c_{\rm s}}}
\newcommand{\diff}{\ensuremath{{\rm d}}}
\newcommand{\e}{\ensuremath{{\rm e}}}
\newcommand{\const}{\ensuremath{{\rm const}}}
\newcommand{\Mach}{\ensuremath{{\cal M}}}
\renewcommand{\vector}[1]{\ensuremath{\bm{#1}}}
\renewcommand{\i}{\ensuremath{\textrm{ i}}}

\newcommand{\cmsqg}{\ensuremath{\rm cm^2\,g^{-1}}}
\newcommand{\km}{\ensuremath{\rm km}}
\newcommand{\keV}{\ensuremath{\rm keV}}
\newcommand{\Msun}{\ensuremath{\rm M_\odot}}
\newcommand{\Msunyr}{\ensuremath{\Msun \rm \, yr^{-1}}}

\newcommand{\ppardir}[2]{\frac{\partial}{\partial #1} \left( #2\right)}
\newcommand{\pardir}[2]{\frac{\partial #2}{\partial #1}}

\newcommand{\alert}[1]{\color{red}#1\color{black}}

\makeatletter
\newcommand{\github}[1]{%
   \href{#1}{\faGithubSquare}%
}
\makeatother

\begin{document}

\title[Simulating accretion spreading layers]{Numerical approach to compressible shallow-water dynamics of neutron-star spreading layers}

\author{Aleksandr Rusakov}
\affiliation{Saint Petersburg State University, 7/9 Universitetskaya nab., St. Petersburg, 199034, Russia; rusakov.sasha@outlook.com} 

\author{Pavel Abolmasov}
\affiliation{The Raymond and Beverly Sackler School of Physics and Astronomy, Tel Aviv University, Tel Aviv 69978, Israel}

\author{Omer Bromberg}
\affiliation{The Raymond and Beverly Sackler School of Physics and Astronomy, Tel Aviv University, Tel Aviv 69978, Israel}

\begin{abstract}
A weakly magnetized neutron star (NS) undergoing disk accretion should release about half of its power in a compact region known as the accretion boundary layer. 
Latitudinal spread of the accreted matter and efficient radiative cooling justify the approach to this flow as a two-dimensional \emph{spreading layer} (SL) on the surface of the star.
Numerical simulations of SLs are challenging because of the curved geometry and supersonic nature of the problem. 
% To study the steady-state structure of the SL and its temporal behavior, we need a numerical technique solving the equations of compressible hydrodynamics on the surface of the sphere, free from singularities and reliable in the supersonic regime up to Mach number of about tens. 
 We develop and test a new two-dimensional hydrodynamics code {\tt SPLASH} that uses the multislope second-order MUSCL scheme in combination with an HLLC+ Riemann solver on an arbitrary irregular mesh on a spherical surface. 
The code is suitable and accurate for Mach numbers up to 5-10. Adding sinks and sources to the conserved variables, we simulate early stages of constant-rate accretion onto a spherical NS. During these stages of accretion, heating in the equatorial region triggers convective instability that causes rapid mixing in the latitudinal direction. One of the outcomes of the instability is the development of a `tennis ball' pattern rotating at a frequency considerably different from the rotation frequency of the matter. In the simulated variability curves, there are signatures of both the pattern rotation and the rotation of the SL, as well as their harmonics. The pattern and cyclonal vortices associated with it are probably a manifestation of an inertial oscillation mode excited within the layer.  \hfill \github{https://github.com/TURBOLOSE/SPLASH}   
 %The code is suitable and accurate for Mach numbers up to 5-10. Adding sinks and sources to the conserved variables, we simulate constant-rate accretion onto a spherical NS. During the early stages of accretion, heating in the equatorial region triggers convective instability that causes rapid mixing in the latitudinal direction. One of the outcomes of the instability is the development of a two-armed `tennis ball' pattern rotating as a rigid body. From the point of view of a high-inclination observer, its contribution to the light curve is seen as a high-quality-factor quasi-periodic oscillation mode with a frequency considerably smaller than the rotation frequency of the matter in the SL. Other variability modes seen in the simulated light curves are probably associated with low-azimuthal-number Rossby waves. \hfill \github{https://github.com/TURBOLOSE/SPLASH}          

\end{abstract}




\section{Introduction} \label{sec:intro}

A typical neutron star (NS) is born in the end of the evolution of a massive star and initially possesses a strong magnetic field and a large angular momentum. 
Ohmic losses in the crust \citep{2018MNRAS.473.2771B} and field burial during accretion \citep{2004MNRAS.351..569P} destroy the magnetic field, creating a population of old, weakly magnetized, rapidly rotating, relatively massive NSs observed as accretors in low-mass X-ray binaries (LMXBs, see \citealt{2023hxga.book..120B}). 

The absence of strong magnetic fields allows LMXBs to have accretion disks extending all the way toward the NS surface (or the innermost stable circular orbit that might be slightly above the surface; this scenario is known as the accretion-gap case, see \citealt{1990ApJ...358..538K}). 
The structure of the innermost parts of the flow, where accreting matter slows down from approximately Keplerian rotation in the disk to the much slower rotation of the NS, is understood in very little detail. 
This part of the flow is called the accretion boundary layer, or simply boundary layer (BL). 

Weakly magnetized NSs in LMXBs are among the brightest and the most numerous X-ray sources in the sky, and their timing and spectral properties have been studied since the 70s.  
The spectral and timing properties of the BL allow for the separation of its contribution to the observed flux from that of the disk \citep{2003A&A...410..217G}. 
With advances in X-ray polarization measurement techniques, timing and spectral data are now complemented by polarization measurements for several sources \citep{2024Galax..12...43U}.
Some of them demonstrate behavior unexpected from LMXBs, such as considerable variations of the polarization angles. 
All this creates a demand for new theoretical work and numerical simulations of the accretion flows onto weakly-magnetized NSs. 

The rich phenomenology of NS LMXBs includes different quasi-periodic oscillations (QPOs), in particular  kilohertz-timescale QPOs, kHz~QPO \citep{2000ARA&A..38..717V}.
Their periods are of the same order as the free-fall time near the surface. 
Besides, the properties of kHz QPOs in NS systems are profoundly different from the short-timescale QPOs in the systems containing black holes \citep{2019NewAR..8501524I}. 
Their observational properties suggest that this specific oscillation mode must be related to the processes near the surface of the accreting star.  
Obviously, some version of a BL is a good candidate for the origin site of any fast variability of a weakly magnetized accreting NS. 

First models of the BL treated it as a part of the disk \citep{1986MNRAS.220..593P, 1988AdSpR...8b.135S} where the Keplerian rotation of the thin disk decreases to the much smaller value matching the spin of the star. 
One of the insights of such an approach is the result that the radial extent of the layer should be small compared to the thickness of the disk, making it essentially a three-dimensional flow with a complicated geometry.
Modeling such a flow must take into account the latitudinal components of the velocities within it, that are usually ignored in accretion disk models. 

An important step in understanding the BL dynamics is the concept of spreading layer (SL), introduced by \citet{1999AstL...25..269I} and further developed in \citet{2006MNRAS.369.2036S} and \citet{2010AstL...36..848I}. 
In this approach, the latitudinal motions dominate over radial, and efficient radiative cooling from the surface makes the flow effectively two-dimensional, except for a narrow injection region near the equator.  
Existing analytic models of the SL are one-dimensional and steady-state. 

One would expect the SL to have two- and three-dimensional instability and variability modes. 
An analytical study of such oscillation modes and their stability was done by \citet{2003MNRAS.342.1156W}. 
An important step in applying these results to real flows and making a self-consistent model of an SL would be two-dimensional numerical simulations. 
An attempt to consider supersonic two-dimensional dynamics on a sphere was done by \citet{2020A&A...638A.142A}, who have also found an inertial oscillation mode potentially responsible for the kHz QPO mentioned above. 
The method used in this paper to address the flow dynamics was by solving the system of equations in terms of spherical harmonics. 
For subsonic motions, this is an extremely reliable technique, usually applied to shallow-water equations \citep{1995JCoPh.119..164J}. 
Deriving the system of compressible shallow-water equations is straightforward.

The real NS SLs are expected to be moderately supersonic, with the rotational Mach number spanning the range between several and several tens in different parts of the flow. 
For the spectral approach, supersonic flows are hard to simulate because of the Gibbs phenomenon \citep{1997SIAMR..39..644G} caused by shock fronts and strong density and pressure contrasts. 
To be able to handle correctly the dynamics of an SL, a hydrodynamic code should work reliably at high Mach numbers. 
Besides, the correct description of the flow close to the poles of the sphere requires that the code should not have a singularity in the polar regions. 
A solution to all these numerical problems is to develop a numerical code for solving compressible shallow-water equations using a mesh without singular points, that is able to handle large Mach numbers and large density and pressure contrasts and discontinuities. 

In this paper, we present a new numerical code {\tt SPLASH} (SPreading LAyer Simulation for Hydrodynamics) that solves the compressible shallow-water equations on a sphere using an arbitrary irregular mesh with the help of a second-order MUSCL scheme. 
The paper is organized as follows. In Section~\ref{sec:phys}, we introduce the main equations and physical quantities.
In Section~\ref{sec:num}, the numerical algorithm is described. 
In Section~\ref{sec:tests}, we give the results of stability tests, and  in Section~\ref{sec:SL} the results of our simulations of an SL on a spherical star. 
We discuss the results in Section~\ref{sec:disc} and make conclusions in Section~\ref{sec:conc}. 

\section{Physical problem}\label{sec:phys}

\subsection{Typical values and basic equations}

We will consider a spherical NS with typical global parameter values: mass $M = 1.5\Msun$ and radius $R = 10\km$. 
Its rotation will be considered negligibly slow, but we focus on the dynamics of an envelope, that may rotate at a considerable fraction of the Keplerian velocity. 
We will consider a fixed mass accretion rate of $\dot M = 10^{-8}\Msunyr$. 
In terms of emitted luminosity, this is typical for a Z-source similar to Sco~X-1 \citep{1989A&A...225...79H}. 
The source of accreting matter has net angular momentum about 80\% of Keplerian and is limited to an equatorial band as one would expect for a source accreting through a disk. 

For a mass accretion rate close to the Eddington limit, the expected effective temperature of the source is $\sim 1\keV$. 
On long enough time scales, the matter in the layer slows down due to some kind of (very weak) interaction with the crust of the NS, cools down and joins the crust. 
The time needed to lose the excess angular momentum may be very long. 
According to \citet{1999AstL...25..269I}, it should be of the order of thousands of Keplerian times. 
The correlation times related to the parallel-tracks phenomenon \citep{2001ApJ...561..943V,2021A&A...647A..45A} suggest even longer times of the order of millions of Keplerian scales. 
Slow dissipation in the SL allows us to estimate its surface density as $\Sigma \sim \dot{M} t_{\rm corr} / 4\uppi R^2 \sim 10^8\unit{\gram\per\square\centi\metre}$, where $t_{\rm corr}$ is the characteristic correlation time scale of the BL (estimated as minutes to hours from the parallel tracks' effect). 
Below, we will consider an initial configuration with a mean surface density of $\sim 10^7\unit{\gram\per\square\centi\metre}$. 
The internal temperature in the layer is $T \sim  100\keV$, implying that radiation pressure is important and matter is mostly ionized (see also estimates in \citealt{1999AstL...25..269I} and \citealt{2020A&A...638A.142A}).  
The speed of sound is $\cs \sim 0.01c$, where $c$ is the speed of light. 
The velocity of the accreting matter reaches about $0.5c$, hence the expected Mach number is of the order of several tens. 

The equations we are solving are vertically integrated equations of ideal hydrodynamics, including the continuity equation
\begin{equation}
    \pardir{t}{\rho} + \nabla \cdot \left( \rho \vector{v}\right) = 0,
\end{equation}
Euler equations
\begin{equation}
    \pardir{t}{\vector v} + (\vector{v}\cdot \nabla) \vector{v} = - \frac{1}{\rho} \nabla P - \vector{g},
\end{equation}
and energy conservation equation
\begin{equation}
    \ppardir{t}{e+\frac{v^2}{2}\rho+\Phi\rho} +  \nabla \cdot \left[ \left( e + P + \frac{v^2}{2}\rho + \Phi \rho\right) \vector{v} + \vector{q}\right] = 0, 
\end{equation}
where $e$ is internal energy density, $P$ is pressure, $\Phi = -GM/(R+h)$ is gravitational potential, $ \vector{g} = -\nabla \Phi$, and $h$ is elevation above the surface. 
We use a spherical coordinate system where the position is given by a radius vector
\begin{equation}
    \vector{r} = (R+h)\left( 
\begin{array}{c}
     \sin \theta \cos \varphi  \\
      \sin \theta \sin \varphi \\
      \cos \theta \\
\end{array}
    \right),
\end{equation}
and $h \ll R$. For the vertically integrated equations, we will only take into account tangential velocity components $v_\theta$ and $v_\varphi$ along the coordinate vectors of the spherical coordinate system. 

We treat these equations as conservation laws of the form
\begin{equation}
    \ppardir{t}{u(\vector{r}, t)} + \nabla \cdot \vector{F}_u (\vector{r}, t) = S_u(\vector{r}, t),
\end{equation}
where $u$ is the density of the particular conserved quantity,  $\vector{F}_u$ is the corresponding flux vector, and $S_u$ is the corresponding source. 
All the quantities are expressed as functions of the position on the sphere $\vector{r}$ and time $t$. 

The conserved quantities used in the calculations are surface density
\begin{equation}
    \Sigma = \int \rho \diff h,
\end{equation}
angular momentum surface density
\begin{equation}
    \vector{l} = \int \rho (\vector{r}\times \vector{v}) \diff h,
\end{equation}
and surface density of energy
\begin{equation}
    E = \int \left( e + \frac{v^2}{2} \rho + \Phi \rho  \right) \diff h.
\end{equation}
The local energy density $e$ is related to pressure $P$ by the equations for a fluid with adiabatic index $\gamma = 4/3$,
\begin{equation}
    e = \frac{P}{\gamma-1}.
\end{equation}
For the integration scheme, we will use the values of the above five vertically integrated quantities in the centers of the cells of an irregular grid (see Section~\ref{sec:num:sph} for details). 
% Below, we introduce the corresponding flux variables, the sources and sinks we use in the code, and the conversion between the primitives, the conserved quantities, and the fluxes.
Mass flux is calculated as 
\begin{equation}\label{E:flux:Sigma}
    \vector{F}_\Sigma = \Sigma \vector{v}.
\end{equation}
For the other quantities, the fluxes and sources are introduced in the following subsections. 

\subsection{Velocity and angular momenta}

A fluid element on the surface of the sphere tends to conserve its angular momentum. 
Gravity directed along the radius vector gives zero contribution to the angular momentum change, as the torque it creates is proportional to $\vector{r} \times \vector{g} = 0$. 
Including surface deformation due to rotation or tangential stresses would introduce force components along the surface that we will ignore in this paper. 
All the changes in angular momentum come from the forces not aligned with the radius vector, and they may be included in the right-hand side of the conservation laws. 
Angular momentum per unit surface area is
\begin{equation}\label{E:cons:li}
    \vector{l} = \Sigma \vector{r} \times  \vector{v}.
\end{equation}
To convert the primitives using the conservatives, we need to divide the mass by the surface area of the cell $K_i$.
To get velocities from $\vector{l}$, one can take a vector product of (\ref{E:cons:li}) by $\vector{r}$
\begin{equation}
    \vector{r} \times \vector{l} = \Sigma \left(  (\vector{r} \cdot \vector{v}) - \vector{v} R^2  \right) = -\Sigma R^2 \vector{v},
\end{equation}
where we neglect the next-order terms in $h/R$, assuming the layer is thin in vertical direction. 
The last step is possible because all the motions are restricted to the surface of the sphere, and thus are orthogonal to the radius vector. 
This means that the angular momentum vector contains all the necessary information to restore the velocity components, and
\begin{equation}
    \vector{v} = - \frac{\vector{r} \times \vector{l}}{\Sigma R^2} .
\end{equation}
Vector $\vector{l}$ is a vector in 3D space, and thus needs to be defined in a global coordinate system unrelated to the coordinates on the sphere. 
We will use the three Cartesian components of the angular momenta.

Angular momentum flux is obtained by integrating the stress $T_{\alpha \beta} = \rho v_\alpha v_\beta + \delta_{\alpha \beta}P$ over the radial coordinate and taking a vector product with $\vector{R}$. 
Here, the Greek indices correspond to the spatial coordinates in the 3D space. 
Tensor formalism is applied to Cartesian 3D coordinates only and thus does not require a distinction between covariant and contravariant components. 
Given $\vector{n}$ is a unit normal vector to a cell edge tangential to the surface of the sphere, the flux of the $\alpha$ component of $\vector{l}$ along the normal is
\begin{equation}\label{E:flux:l}
\vector{F}_{l, \alpha} =  l_\alpha  \vector{v} -  \left(\vector{n} \times \vector{R}\right)_\alpha \Pi.
\end{equation}

\subsection{Vertical structure and energy}\label{sec:phys:vert}

In the absence of vertical velocities, the vertical component of the Euler equation becomes
\begin{equation}
    \frac{\diff P}{dh} = - g_{\rm eff} \rho,
\end{equation}
where 
\begin{equation}\label{E:geff}
    g_{\rm eff} = \frac{GM}{R^2} - \frac{v^2}{R}.
\end{equation}
Velocity $v$ here has two tangential components, $v_\theta$ (along the polar angle direction) and $v_\varphi$ (azimuthal). 
If we assume that, for given coordinates on the sphere $\theta$ and $\varphi$, $P \propto \rho^{\gamma_{\rm v}}$ with changing $h$. 
% We will assume this exponent is the same as the global adiabatic index, $\gamma_{\rm v} = \gamma$, but it is not necessarily the case in general.
The value of $\gamma_{\rm v}$ does not necessarily coincide with the actual adiabatic index $\gamma$ and does not enter the final vertically integrated equations.

If the vertical structure is affected by radiation transfer, $\gamma_{\rm v} \simeq 4/3$.
Vertical integration results in
\begin{equation}
    \rho = \rho_0 \left( 1 - \frac{h}{H}\right)^{\frac{1}{\gamma_{\rm v}-1}},
\end{equation}
\begin{equation}
    P = P_0 \left( 1 - \frac{h}{H} \right)^{\frac{\gamma_{\rm v}}{\gamma_{\rm v}-1}},
\end{equation}
where $H$ is the radius where density and pressure become zero,
\begin{equation}
    H = \frac{\gamma_{\rm v}}{\gamma_{\rm v}-1} \frac{P_0}{g_{\rm eff}\rho_0}.
\end{equation}
By direct integration, one can prove that
\begin{equation}
    P_0 = g_{\rm eff} \Sigma
\end{equation}
and 
\begin{equation}
    H = \frac{2\gamma_{\rm v}-1}{\gamma_{\rm v}-1} \frac{\Pi}{g_{\rm eff} \Sigma}.
\end{equation}
Total surface energy density 
\begin{equation}
\begin{split}
    E = \int_0^{H} \left( \frac{P}{\gamma-1} + \rho \Phi + \rho \frac{v^2}{2}\right) \diff h =\\ 
    = \frac{\gamma}{\gamma-1} \Pi + \Sigma \frac{v^2}{2} + g_{\rm eff} \Sigma \Delta R, 
\end{split}
\end{equation}
where 
\begin{equation}
    \Phi = g h + \Delta \Phi,
\end{equation}
and $\Delta R$ is the deformation of the stellar surface related to rotation (the term is analogous to $\Delta \Phi$ used in \citealt{2020A&A...638A.142A}).
If we want the deformation to correspond to a certain equilibrium rotation figure, we can replace in the first order in $\Delta R$, 
\begin{equation}
    \frac{GM}{R_*^2}\Delta R \simeq -v_0^2/2 = -\frac{1}{2}\Omega_0^2 R_*^2 \sin^2\theta.
\end{equation}
In such an approximation, total energy is
\begin{equation}\label{E:cons:E}
    E   = \frac{\gamma}{\gamma-1} \Pi + \frac{1}{2} \Sigma \left( v_\theta^2 + v_\varphi^2 - \Omega_0^2 R_*^2\sin^2\theta\right) .
\end{equation}
In this paper, we restrict ourselves to the case of a spherical star, setting $\Omega_0 = 0$. 
Energy flux is found as
\begin{equation}\label{E:flux:E}
\begin{split}
\displaystyle    \vector{F}_E = \int_0^{H} \left( \frac{\gamma P}{\gamma-1}  + \rho \Phi + \rho \frac{v^2}{2}\right) \vector{v} \diff r = \\
 = \displaystyle  \left[ \frac{2\gamma-1}{\gamma-1} \Pi +  \frac{1}{2} \left( v_\theta^2 + v_\varphi^2 - \Omega_0^2 R_*^2\sin^2\theta\right) \Sigma \right] \vector{v} .
\end{split}
\end{equation}
Thermal contribution to vertically integrated energy density contains coefficients that differ from their three-dimensional analogues. 
However, defining two-dimensional effective adiabatic exponent $\Gamma = 2 - 1/\gamma$, we can restore the original form of the expressions 
\begin{equation}\label{E:cons:E2d}
    E   = \frac{1}{\Gamma-1} \Pi + \frac{1}{2} \Sigma \left( v_\theta^2 + v_\varphi^2 - \Omega_0^2 R_*^2\sin^2\theta\right) ,
\end{equation}
\begin{equation}\label{E:flux:E2d}
% \begin{array}{l}
\displaystyle    \vector{F}_E =  \left[ \frac{\Gamma}{\Gamma-1} \Pi +  \frac{1}{2} \left( v_\theta^2 + v_\varphi^2 - \Omega_0^2 R_*^2\sin^2\theta\right) \Sigma \right] \vector{v} .
%     \end{array}
\end{equation}
As a consequence, adiabatic relationship between vertically integrated pressure and density is $\Pi \propto \Sigma^\Gamma$. 
Hence, the speed of sound is also $\cs = \sqrt{\Gamma \Pi/\Sigma}$. 

\subsection{The case of gas and radiation pressure}

We assume the modeled matter to be a mixture of gas and radiation. We introduce the parameter $\beta$ as the ratio
of gas pressure to the total pressure of gas and radiation. 
In a real SL, most of the energy is released near its bottom and the vertical structure is formed by radiation transfer. 
In the assumption of constant opacity $\varkappa = $ const, this implies $\beta$ independent of the vertical coordinate and effective $\gamma_{\rm v} = 4/3$ (see \citealt[section 2.2]{1999AstL...25..269I}).
The effective adiabatic index present in equations (\ref{E:cons:E}) and (\ref{E:flux:E}) is effective three-dimensional $\gamma$, that may be proven by direct integration over $h$. 
Its dependence on $\beta$ is set by 
\begin{equation}
    u = 3\left( 1- \frac{\beta}{2}\right) p = \frac{p}{\gamma-1}, 
\end{equation}
that implies
\begin{equation}
    \gamma = \frac{8 - 3\beta}{3(2-\beta)}.
\end{equation}
The corresponding two-dimensional adiabatic index is 
\begin{equation}
    \Gamma = \frac{10-3\beta}{8-3\beta}.
\end{equation}
Here, $\beta$ may be computed by solving equation (20) from \cite{2020A&A...638A.142A}
\begin{equation}\label{E:beta}
\frac{\beta }{(1-\beta)^{1/4}}=\frac{4}{5}\frac{k}{m} \left( \frac{3}{4} \frac{c}{\sigma_{\rm SB}}g_{\rm eff} \Sigma \right)^{1/4} \frac{\Sigma}{\Pi}.
\end{equation}
Here, $m$ is the mean mass of a massive particle $m\approx0.6m_p$ $k$ is the Boltzmann constant, and  $\sigma_{\rm SB}$ is the Stefan–Boltzmann constant.

An approximation can be made for the left-hand side of equation~(\ref{E:beta})
\begin{equation}
      % \frac{\beta}{(1-\beta)^{1/4}(1-\beta/2)} \approx \begin{cases} 
      %      \frac{\beta}{1-\beta} &   \beta \leq \beta_{\rm switch} \\
      %     \frac{2}{\sqrt[4]{1-\beta}}& \beta > \beta_{\rm switch}, 
            \frac{\beta}{(1-\beta)^{1/4}} \approx \begin{cases} 
           \frac{\beta}{1-\beta} &   \beta \leq \beta_{\rm switch} \\
          \frac{1}{(1-\beta)^{1/4}}& \beta > \beta_{\rm switch}, 
\end{cases}
\end{equation}
where $\beta_{\rm switch} \simeq 0.5497$. 

The precise value of $\beta_{\rm switch}$ is chosen to match the values given by the two approximations given above. After finding the approximation for $\beta$, we perform 2 Newton iterations and achieve the accuracy of $\approx 10^{-6}$.

\begin{comment}
\alert{We already defined $\gamma$ as $\gamma_{\rm local}$; I would rather suggest to modify expressions for energy and energy flux in order to account for gas and radiation, the gas having $\Gamma = 7/5$ and radiation $\Gamma=5/4$; $\beta$ is restored using the expressions above and used to recover the pressure ratio. Shiokawa was just an example.}
Using expression from \cite{Shiokawa_2015} we can find local $\gamma$ as
\begin{equation}
    \gamma_{\rm local} = \gamma - \frac{\gamma -4/3}{1+u_{\rm gas}/u_{\rm rad}},
\end{equation}
where $u_{\rm gas}$ and $u_{\rm rad}$ are internal energies.
We can derive their ratio using $\beta$:
\begin{equation}
    u_{\rm gas}/u_{\rm rad} = \frac{\beta}{1-\beta} \frac{1}{3(\gamma-1)}.
\end{equation}

In this way, we compute $\gamma$ and $\Gamma = 2-1/\gamma$ to correctly account for the state of matter at any moment.
\end{comment}

\subsection{Sources and sinks}\label{sec:phys:src}

%With the corrections to the potential, there is no need for energy source, and the source of angular momentum (per unit surface area) may be written as 
The deformation of the surface of the NS is included in the energy equation through the potential terms in equations~(\ref{E:cons:E2d}) and (\ref{E:flux:E2d}), therefore there is no need to include the associated work as a source of energy. 
%The deformation of the NS surface is included in the energy flux and does not contribute the energy source. 
However, the deformation should be included in the angular momentum source as
\begin{equation}
    \dot{\vector{l}}_{\Phi} = \Sigma \vector{r} \times \vector{g} = 
    \vector{e}_\theta \times \vector{r}\  \Sigma \Omega_0^2 R_* \sin \theta \cos \theta.
\end{equation}
In order to properly simulate accretion and other physical effects relevant for the SL problem, we include additional source and sink terms.
Accretion is included as a mass source dependent on the coordinates as 
\begin{equation}\label{E:source:Gauss}
  \displaystyle  \dot{\Sigma}_{\rm acc} = \frac{\dot{M}}{4\uppi R^2 } \frac{1}{\sqrt{2\pi} \sigma_\alpha} \e^{- \frac{\cos^2\alpha}{2\sigma_\alpha^2}},
\end{equation}
where $\alpha$ is the polar angle with the axis of the disk that we assume here to be inclined by $\alpha_{\rm tilt} = 6^\circ$ with respect to the rotation axis of the NS, $\sigma_\alpha$ determines the width of the band within which accretion occurs, and 
\begin{align}
    \cos \alpha &= \frac{\vector{r} \cdot \vector{\Omega}_{\rm orb}}{|\vector{\Omega}_{\rm orb}|}
         = \cos \theta \cos \alpha_{\rm tilt} + \sin\theta \sin \alpha_{\rm tilt} \sin \varphi,
\end{align}
where $\vector{\Omega}_{\rm orb}$ is the orbital rotation frequency of the accreting matter. 
This mass source distribution and the associated injection velocity profile (see below) correspond to a thin accretion disk with the relative thickness $\sim 10\%$. 
% We chose $\sigma_\alpha$ to create a distribution with $\rm FWHM=0.1 \rm R$, making the accretion disk geometrically thin. 
% The most important source term is the mass accretion term $\dot M$ that is turned into a density source $\dot \Sigma_{\rm acc}$. The accretion happens in a spherical segment with $h=0.1R$ that has a tilt of 6\textdegree.

% We assume the accreted matter has innate angular momentum and energy (per unit area) attached to it. These manifest as follows:
We assume that the net angular momentum and energy content of the accreting matter are set by the conditions in the disk, that allows to set the sources as

\begin{equation}
%  \begin{split}
    \dot{\vector{l}}_{\rm acc} = \dot \Sigma_{\rm acc} \left( \vector{r} \times  \vector{v}_{\rm orb}\right) 
\end{equation} 
 and   
\begin{equation}
\dot{E}_{\rm acc}= \dot \Sigma_{\rm acc} \left( \left( \frac{E}{\Sigma}\right)_{\rm d}+\frac{1}{2}(\vector{v}_{\rm orb}-\vector{v})^2\right).
%     \end{split}
\end{equation}
We assume $\vector{v}_{\rm orb} = \vector{r} \times \vector{\Omega}_{\rm orb}$ to be slightly lower than the Keplerian velocity at the surface. We consider $\vector{\Omega}_{\rm orb}$ to be constant.
% and have a negligible radial velocity.
% This means we can treat it as a rotating motion around the same 6\textdegree \ tilted axis.

The first sink term we introduce is the loss of mass from the bottom of the spreading layer. % While the accretion only happens in a thin line on the surface, the mass loss occurs everywhere on a sphere and it is proportional to the local density $\Sigma$, or
Mass sink is locally proportional to surface density as
\begin{equation}
\dot \Sigma_{\rm fall} = - \frac{1}{t_{\rm fall}} \Sigma.
\end{equation}
We define the constant $t_{\rm fall}$ from the assumption of mass balance in the spreading layer. 
It has a physical meaning of the e-folding time for the surface density affected by the sink term only
\begin{equation}
 t_{\rm fall}=\frac{M_{\rm tot} }{\dot  M},
\end{equation}
where $M_{\rm tot}$ is the initial mass of the SL. 
% total mass residing in the spreading layer in the initial step. 

If mass is lost from the bottom of the layer at a rate $\dot \Sigma$ (for mass surface density), this loss should be accompanied with the loss in angular momentum and energy, proportional to $\dot \Sigma$. 
% If $\vector{l}$ is the angular momentum per unit area,
Associated momentum loss is 
\begin{equation}
\dot{\vector{l}}_{\rm fall} = \frac{\vector{l}}{\Sigma} \dot \Sigma_{\rm fall} ,
\end{equation}
meaning that the net angular momentum of the matter locally lost from the system is the local vertically average value. 
% meaning that the matter lost from the system has average net angular momentum, if there is no tangential stress on the boundary 
The amount of energy lost from the cell is defined by the vertical energy flux \citep[section I.6]{landafshitz}
\begin{equation}
    F_{E, r} = v_r \rho \left( \frac{v^2}{2} + \frac{\gamma_{\rm v}}{\gamma_{\rm v}-1} \frac{P_0}{\rho_0} \right).
\end{equation}
We do not know the vertical velocity $v_r \ll v$, but $\rho v_r = \dot \Sigma_{\rm fall}$ (mass flux through the surface), that allows (taking into account the results of Section~\ref{sec:phys:vert}) to re-write the expression as
\begin{equation}
\dot E_{\rm fall} = \dot \Sigma_{\rm fall} \left( \frac{v^2}{2} + \frac{\Gamma}{\Gamma-1} \frac{\Pi}{\Sigma} \right).
\end{equation}
Another sink term we introduce is the radiation energy loss in a way it is described in \cite{2020A&A...638A.142A}:
\begin{equation}\label{E:sink:rad}
\dot E_{\rm rad} = -\frac{c g_{ \rm eff}}{\varkappa} \left( 1-\beta \right).
\end{equation}
Here, $\varkappa$ is the Rosseland average opacity (we assume Thomson scattering the main opacity source, that implies $\varkappa \simeq \varkappa_{\rm T} \simeq 0.34 \, \cmsqg$), and $g_{\rm eff}$ is given by equation~(\ref{E:geff}).
%, and
%\begin{equation}
%    g_{\rm eff} = - \frac{GM}{R} + \frac{v_\theta^2+v_\varphi^2}{2}
%\end{equation}
%is the effective free fall acceleration at the given point, 

% Precise value can be determined both computationally or analytically to ensure continuity. 
% This piecewise approximation yields decent results in most of the use cases and allows to avoid computationally expensive solution.

% The total of all sources and sinks for each conservative variable:
The general expressions for the source terms we use are
\begin{equation}\label{E:sources}
\begin{split}
\dot \Sigma &= \dot\Sigma_{\rm acc}+ \dot\Sigma_{\rm fall},\\
\dot{\vector{l}} &= \dot{\vector{l}}_{\Phi} + \dot{\vector{l}}_{\rm acc} + \dot{\vector{l}}_{\rm fall},\\ 
\dot E &= \dot E_{\rm acc} + \dot E_{\rm fall} + \dot E_{\rm rad},
\end{split}
\end{equation}
where the individual source/sink terms are described by the equations given earlier this section.

\section{Numerical algorithm}\label{sec:num}

\subsection{MUSCL scheme for general unstructured meshes}\label{sec:MUSCL}

\begin{figure*}
\begin{center}
\includegraphics[width=1.0\textwidth]{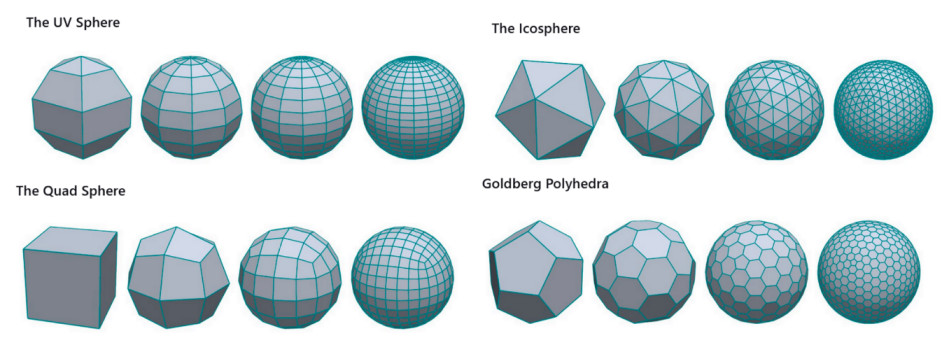}
\end{center}
\caption{Different meshes used from \cite{sieger_2021_generating}.}\label{fig:meshes}
\end{figure*}

% The basis for the conservative scheme was 
The conservative scheme we present here is based on the multislope MUSCL (Monotonic Upstream-centered Scheme for Conservation Laws \citealt{VANLEER1979101}) method for general unstructured meshes described in \citet{cletouze_2015_multislope}. The main advantage of this scheme is its second order of error, that does not come at the cost of stability. It demonstrates lower accuracy compared to the ENO and WENO  \citep{Shu_2020}; however, it is possible to implement on an unstructured mesh. 
We adapt the planar 2D method for use on an arbitrary spherical mesh. We tested several different meshes (Figure \ref{fig:meshes}) in order to find which of them worked best for the given problem.

MUSCL scheme works for any hyperbolic conservation equation system with initial condition.
The general form for each of the equations of the system is 
\begin{equation}
	\label{eq:sys}
	\partial_t u(\vector{x},t) + \vector{\nabla} \vector{F} (\vector{x},t) = \dot{u} \quad
    u(\vector{x},t=0)=u_0,
\end{equation}
where $u = \Sigma$, $\vector{l}$, and $E$, fluxes are calculated using equations~(\ref{E:flux:Sigma}), (\ref{E:flux:l}), and (\ref{E:flux:E}), and the right-hand side contains the sources given by equation~(\ref{E:sources}). 

As we are dealing with a sphere, the boundary condition is not necessary. Let $\Omega \subset {\cal R}^2$ be a domain, $x \in \Omega$. Let us look at a discretization of $\Omega$ made up of polygons $K_i$. We can define $\mathcal{V}(i)$ as the
vicinity of the element $K_i$, defined as the set of neighboring elements $K_j$ that share with $K_i$ a common edge $S_{ij} = K_i \cap K_j$. $U_i^{n}$ defines the discrete value of a conservative variable $u$ in the center of the face $K_i$ at the moment $t_n=t_0+n\Delta t$. Then the numeric scheme can be written as 
\begin{equation}
	\label{eq:sch}
    U_i^{n+1}=U_i^{n}-\Delta t \sum_{\mathcal{V}(i)} \frac{S_{ij}}{K_i} \phi(U_{ij}^n, U_{ji}^n, \vector{n}_{ij},\vector{F}),
\end{equation}
where $S_{ij}$ is the length of the edge between $K_i$ and $K_j$, $K_i$ is the surface area of a cell, $\phi$ is the Riemann solver that is dependent on the physical flux $\vector{F}$, $U_{ij}^n, U_{ji}^n$ are the reconstructed values, $\vector{n}_{ij}$ is the unit normal vector of the edge directed outwards.

\begin{figure}[h]
\begin{center}
\includegraphics[width=0.85\columnwidth]{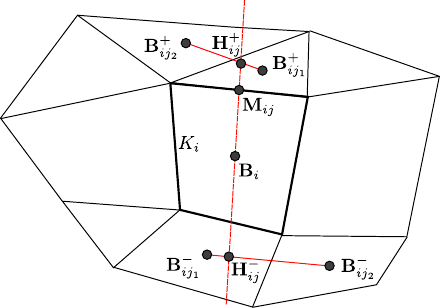}
\end{center}
\caption{Geometric reconstruction for 2D configuration.}\label{fig:geo}
\end{figure}

In order to find a reconstructed value of a conservative variable at the edge center $\vector{M}_{ij}$ (see Figure \ref{fig:geo}) we need to find forward and backward slopes $p_{ij}^+$ and $p_{ij}^-$. With $r=p_{ij}^-/p_{ij}^+$ we can introduce a flux limiter function $\varphi(r)$ that turns two slopes into one limited slope in order to avoid spurious oscillations. The flux limiter should also be dependent on the local geometry. Dropping the temporal index $n$, the reconstructed values can be defined as
\begin{equation}
	\label{eq:recon}
    U_{ij}=U_i+p_{ij}^+ \varphi(r) ||\vector{B}_i \vector{M}_{ij}||.
\end{equation}
We can find slopes by finding two points on our mesh: $\vector{H}_{ij}^+$ and $\vector{H}_{ij}^-$. If we know the values of the conservative variables in this point, we can find the slopes using
\begin{equation} 
\label{eq:slopes}
p_{ij}^+=\frac{U_{H_{ij}^+}-U_i}{||\vector{B}_i \vector{H}_{ij}^+||}, \quad p_{ij}^-=\frac{U_i-U_{H_{ij}^-}}{||\vector{B}_i \vector{H}_{ij}^-||}.
\end{equation}
The process of finding these special points depends on the mesh. An example is given in Figure \ref{fig:geo}. Let us denote the set of all neighboring faces of a cell $K_i$ as $\mathcal{W}(i)$. It is different from $\mathcal{V}(i)$ because $\mathcal{W}(i)$ also contains neighboring faces of $K_i$ that share only one common vertice. 
Connecting the face center $\vector{B}_i$ with the point $\vector{M}_{ij}$ results in a line. We have to find $K_{ij_1}^-$, the most backward neighboring element of $K_i$ (in Figure \ref{fig:geo} it is a face with a center $\vector{B}_{ij_1}^-$). We can define it formally as
\begin{equation} 
\label{eq:bckwrd1}
\cos(\vector{B}_{ij_1}^-\vector{B}_i,\vector{B}_i\vector{M}_{ij}) = \min_{k \in \mathcal{W}(i)}  \cos(\vector{B}_{k}\vector{B}_i,\vector{B}_i\vector{M}_{ij})  .
\end{equation}
Then we also have to define $K_{ij_2}$, the second most backward neighboring element of $K_i$ that is located on the other side of the line $\vector{B}_i \vector{M}_{ij}$. Formally it can be defined as 
\begin{equation} 
\label{eq:bckwrd2}
\cos(\vector{B}_{ij_2}^-\vector{B}_i,\vector{B}_i\vector{M}_{ij}) = \min_{k \in \overline{\mathcal{W}}(i)}  \cos(\vector{B}_{k}\vector{B}_i,\vector{B}_i\vector{M}_{ij})  ,
\end{equation}
where 
\begin{equation} 
\label{eq:Wi}
\begin{split}
\overline{\mathcal{W}}(i) = \{ k \in \{\mathcal{W}(i) \setminus K_{ij_1}^-  \} | \sin(\vector{B}_{k}\vector{B}_i,\vector{B}_i\vector{M}_{ij}^{\perp}) \cdot \\ \cdot \sin(\vector{B}_{ij_1}^-\vector{B}_i,\vector{B}_i\vector{M}_{ij}^{\perp}) \leq 0    \}  .
\end{split}
\end{equation}

Using a similar algorithm, we can find $\vector{B}_{ij_1}^+$ and $\vector{B}_{ij_2}^+$. Then, by finding the intersections between lines $\vector{B}_{i} \vector{M}_{ij}$ and $\vector{B}_{ij_1}^+ \vector{B}_{ij_2}^+$ or $\vector{B}_{ij_1}^- \vector{B}_{ij_2}^-$, we are able to find points $\vector{H}_{ij}^+$ and $\vector{H}_{ij}^-$. 
Finally, in order to find our conservative values in these points, we have to introduce barycentric coordinates
\begin{equation} 
\label{eq:betas_minus}
 \beta_{ij_1}^-=\frac{|| {B}_{ij_2}^- {H}_{ij}^-||}{||{B}_{ij_1}^-{B}_{ij_2}^-||}, \quad \beta_{ij_2}^-=\frac{|| {B}_{ij_1}^- {H}_{ij}^-||}{||{B}_{ij_1}^-{B}_{ij_2}^-||},
\end{equation}
\begin{equation} 
\label{eq:betas_plus}
 \beta_{ij_1}^+=\frac{|| {B}_{ij_2}^+ {H}_{ij}^+||}{||{B}_{ij_1}^+{B}_{ij_2}^+||}, \quad \beta_{ij_2}^+=\frac{|| {B}_{ij_1}^+ {H}_{ij}^+||}{||{B}_{ij_1}^+{B}_{ij_2}^+||}.
\end{equation}
Then the variable values at points $\vector{H}_{ij}^+$ and $\vector{H}_{ij}^-$ can be found as
\begin{equation} 
\label{eq:variables_Hij}
U_{H_{ij}^-} =  \beta_{ij_1}^- U_{ij_1}^- + \beta_{ij_2}^- U_{ij_2}^- , \quad U_{H_{ij}^+} =  \beta_{ij_1}^+ U_{ij_1}^+ + \beta_{ij_2}^+ U_{ij_2}^+ .
\end{equation}

\subsection{Spherical adaptation and meshes}\label{sec:num:sph}

We use meshes created using an algorithm from \cite{sieger_2021_generating} shown in Figure \ref{fig:meshes}. Further, we will use cubic, triangular, and hexagonal mesh. We abstain from using UV-mesh, as it is less symmetrical than the other options; moreover, its polar area is not optimal for the reconstruction algorithm due to having a large number of faces sharing the same vertex.

Cubic, triangular, and hexagonal meshes are based on cube, icosahedron, and dodecahedron, respectively.
The refinement of cubic and triangular meshes is achieved by subdividing each face into 4 parts and then projecting the resulting vertices onto the sphere. This results in a fast and consistent way of creating a fine mesh. One issue with this approach is the impossibility to directly set the desired number of faces. The hexagonal mesh is generated by projecting the centers of a triangular mesh onto a sphere and using them as new vertices.

Different types of mesh have different defining traits. The triangular mesh has a large number of similarly sized faces, so it is more computationally expensive. The cubic mesh has fewer faces on the same level of subdivision; however, the shape and size of these faces differ. In the hexagonal mesh every two neighboring cells share an edge, although some of the faces are pentagons that are considerably smaller than their neighboring faces.

\begin{figure*} % [h]
\begin{center}
\includegraphics[width=0.7\textwidth]{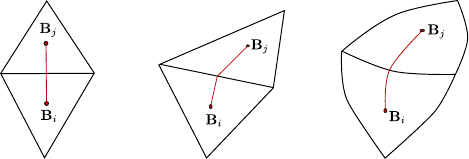}
\end{center}
\caption{Difference between a 2D mesh, a spherical polyhedron mesh and a spherical tessellation.}\label{fig:pln_vs_sph}
\end{figure*}

In order to apply the algorithm to a spherical mesh, several adaptations have to be made. We can treat meshes either like a polyhedra or as spherical tessellations, as shown Fig.~\ref{fig:pln_vs_sph}. The first case requires us to work with broken straight lines, while the second one turns straight lines into great circle arcs and requires spherical trigonometry. After multiple tests, we decided to use spherical tessellation as it is more physically correct and no `mountains' or `trenches' are introduced into geometry. Use of spherical tessellation allows us to use the algorithm from Section ~\ref{sec:MUSCL} with minimal changes, such as replacing lines with great circle arcs and using spherical trigonometry.

Every face is treated as a spherical polygon with vertices $\{\vector{r}_k\}_{k=1}^n$. The center of a cell $\vector{R}_i$ is therefore defined as
\begin{equation}
    \vector{R}_i=\frac{\displaystyle \sum_{k=1}^n \vector{r}_k}{ \Bigg |  \displaystyle \sum_{k=1}^n \vector{r}_k \Bigg | }.
    \label{eq:face_center}
\end{equation}
$\vector{r}_k$ are located on the same unit sphere as $\vector{R}_i$.

One of the main differences lies in the choice of edge normals that are needed for a Riemann solver $\phi(U_{ij}^n, U_{ji}^n, \vector{n}_{ij}, \vector{F})$. As we were going to reconstruct the velocities on the edges, we need to ensure that the velocity vector is orthogonal to the edge then $(\vector{n}_{ij} \cdot \vector{v}) = ||\vector{v}||$. The edge normals are defined as follows: if $\vector{r}_1,\vector{r}_2$ are vertices that define an edge, then 
\begin{equation}
   \vector{n}_{ij}= \frac{(\vector{r}_1+\vector{r}_2) \times (\vector{r}_1-\vector{r}_2)}{||(\vector{r}_1+\vector{r}_2) \times (\vector{r}_1-\vector{r}_2)||} . 
\end{equation}
This gives a normal unit vector that is tangent to the surface of the sphere in the center of the edge. 

\subsection{Choice of the Riemann solver and limiter}

One of the biggest challenges of the problem we are solving is the presence of large velocity differences. 
% significant differences in speed in different regions. 
Because our initial state includes rigid-body rotation, the velocity around the poles is close to zero. On the other hand, the accreted matter that falls onto the surface as a source term can reach Mach numbers $\Mach \sim 10$ and higher, up to $50$ (see Section~\ref{sec:phys}). 
Being able to solve the Riemann problem in both cases required us to choose an all-speed Riemann solver. We chose HLLC+ from \citet{doi:10.1137/18M119032X}, that is based on the original approximate HLLC (Harten--Lax--van Leer Riemann solver with contact restoration) and includes multiple wave fixes for both low and high Mach numbers. We also tested the regular HLLC Riemann solver, as well as the HLLE for isothermal system of equations.

We introduce several changes into HLLC+ in order to adapt it to our choice of conservative variables. HLLC-based Riemann solvers use notations of $\vector{F}(\vector{U}) = v \vector{U}+p \vector{D}$. In our case $U$ is a vector of conservative variables, $v= \vector{v} \cdot \vector{n}$. Replacing $p$ with $\Pi$ we can set $\vector{D}=\left[0,- (\vector{n} \times \vector{R}), S_* \right]$, where $S_*$ is the speed of the middle wave in HLLC Riemann solver. Additionally, we replace normals $\vector{n}$ with $- (\vector{n} \times \vector{R})$ in angular momentum fixes that were introduced in HLLC+.

As for the limiter, we used the hybrid limiter introduced in \cite{cletouze_2015_multislope}. It is proven to be stable for multislope MUSCL reconstruction, and it was the one that worked best in most of the tests.

\subsection{Time-step stability conditions and temporal integration}

In \citet{cletouze_2015_multislope}, the authors provide sufficient stability condition and time-step limitations. Unfortunately, this condition results in unreasonably small $\Delta t$, which required us to choose a time step from other considerations. The sufficient condition allows larger time steps without making the scheme unstable.
We choose a CFL-based time step that accounts for the highest speed on the mesh in a given snapshot.
If $h_0$ is the length of the shortest edge,
\begin{equation}
    \Delta t = \frac{h_0 \cdot \rm{CFL} }{ \max \left( \displaystyle \max_{i \in \Omega} |\vector{v_i}|,  \displaystyle \max_{i \in \Omega} c_i \right)}.
\label{eq:dt}
\end{equation}
After series of different tests, the $\rm CFL$ value of $0.3$ proved to provide stability in most cases and was used to generate all of the test results presented further.

As for the temporal integration, we use explicit second-order Runge-Kutta integrator. Since the reconstruction algorithm has a second order of numerical error, we needed temporal integrator with a corresponding order or error.

\section{Code tests}\label{sec:tests}

The algorithms described above were implemented in the C++ code {\tt SPLASH}, which is publicly available on GitHub \footnote{\url{https://github.com/TURBOLOSE/SPLASH`}}.
% Based on the algorithm described above, a numerical code has been written in C++ programming language. It is publicly available in the online repository \footnote{\url{https://github.com/TURBOLOSE/SPLASH`}}.
The code supports OpenMP parallelism on a single node.

In order to ensure the robustness of the algorithm as well as to test its capabilities, a set of tests has been conducted. These tests were selected in a way that allows to confirm the applicability of the code to different scenarios that may occur in real physical problems. Below is a list of the results of applying the code to different physical problems (steady-state and dynamic).

 In this section and in the following Section~\ref{sec:SL}, we will present two-dimensional maps of vertically-integrated pressure and density in Gall–Peters cylindrical equal-area projection from \cite{gall1885use} defined as
\begin{equation}
\label{eq:proj}
x=\frac{\varphi}{\sqrt{2}}, \quad y=\sqrt{2}\cos \theta.
\end{equation}

\subsection{Stationary rotating atmosphere solution and mesh comparison}\label{sec:res:stat}

With our ultimate goal being the steady-state spreading layer model, we had to test the ability of our code to model stationary solutions.

% Two stationary analytical solutions are described in Appendix~\ref{app:A}. 
The reconstruction scheme allows for the preservation of these profiles with minimal, yet still non-vanishing numerical error. 
The application of Riemann solver ensures that the expected analytical solution is recovered even with high Mach numbers of the flow.

The first test for the family of analytic steady-state solutions for a polytropic rigidly rotating envelope described in Appendix~\ref{app:A}. The adiabatic index $\gamma$ is constant and set to be equal to $4/3$.

\begin{figure*}% [h]
\begin{center}
\includegraphics[width=0.99\textwidth]{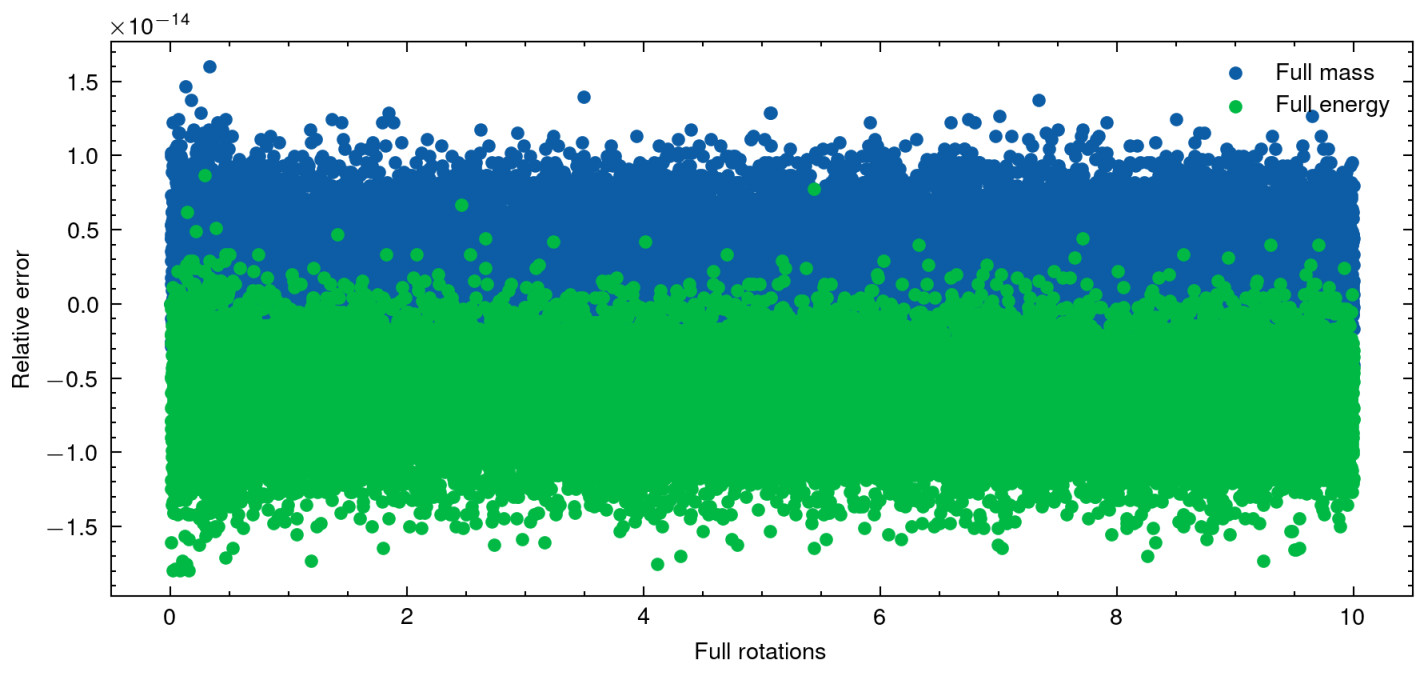}
\end{center}
\caption{Relative error in conservation of total mass and total energy.}\label{fig:err_const_entr}
\end{figure*}

\begin{figure*} %[h]
\begin{center}
\includegraphics[width=0.7\textwidth]{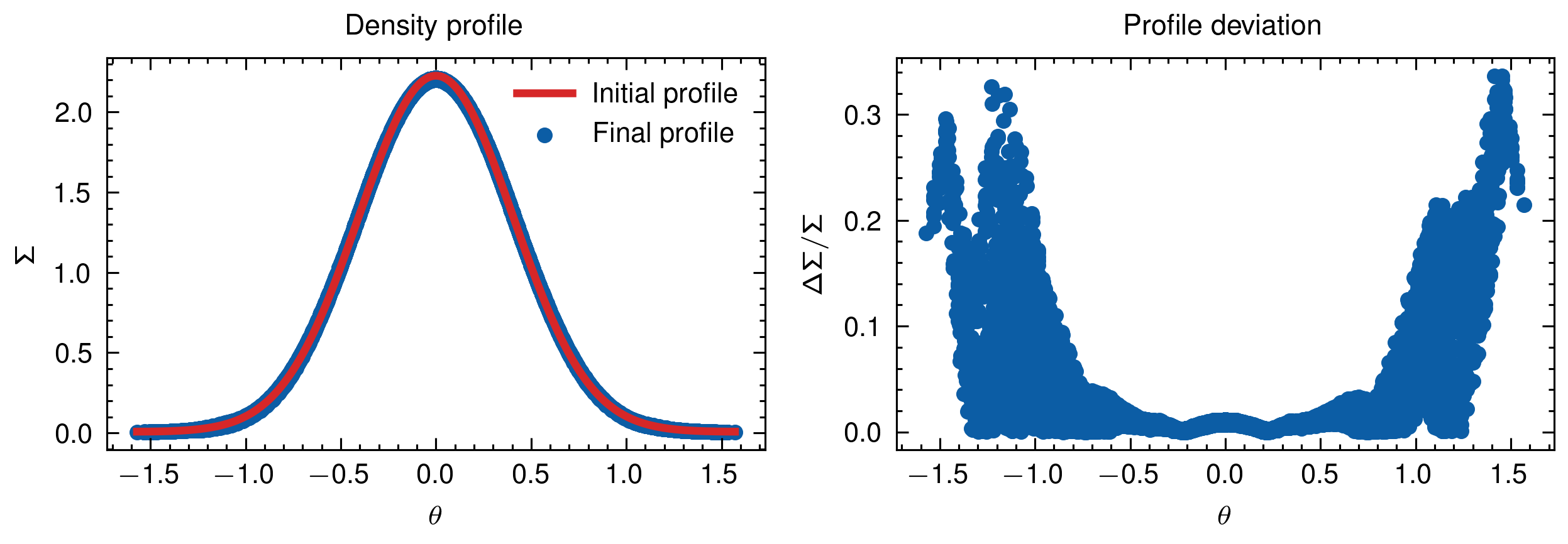}
\includegraphics[width=0.7\textwidth]{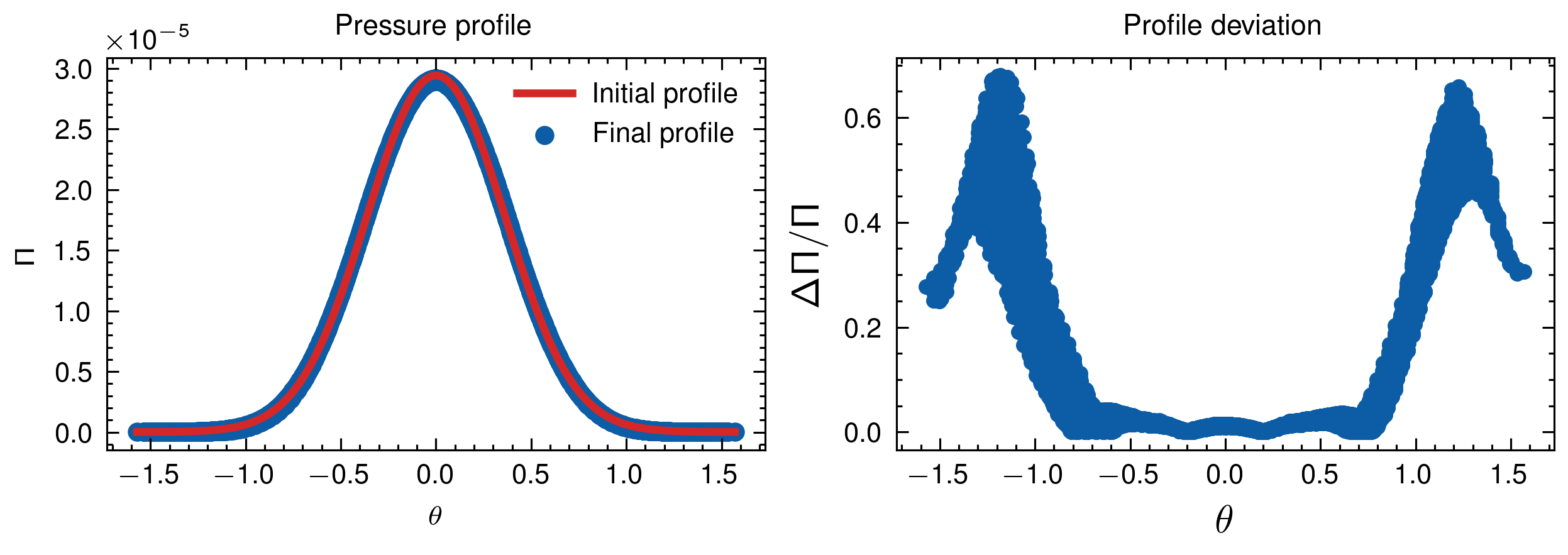}
\end{center}
\caption{Conservation of $\Sigma$ and $\Pi$ profiles after 10 full rotations on a hexagonal mesh with 10242 faces.}\label{fig:sigma_pi_prof_hex}
\end{figure*}

\begin{figure*} %[h]
\begin{center}
\includegraphics[width=0.7\textwidth]{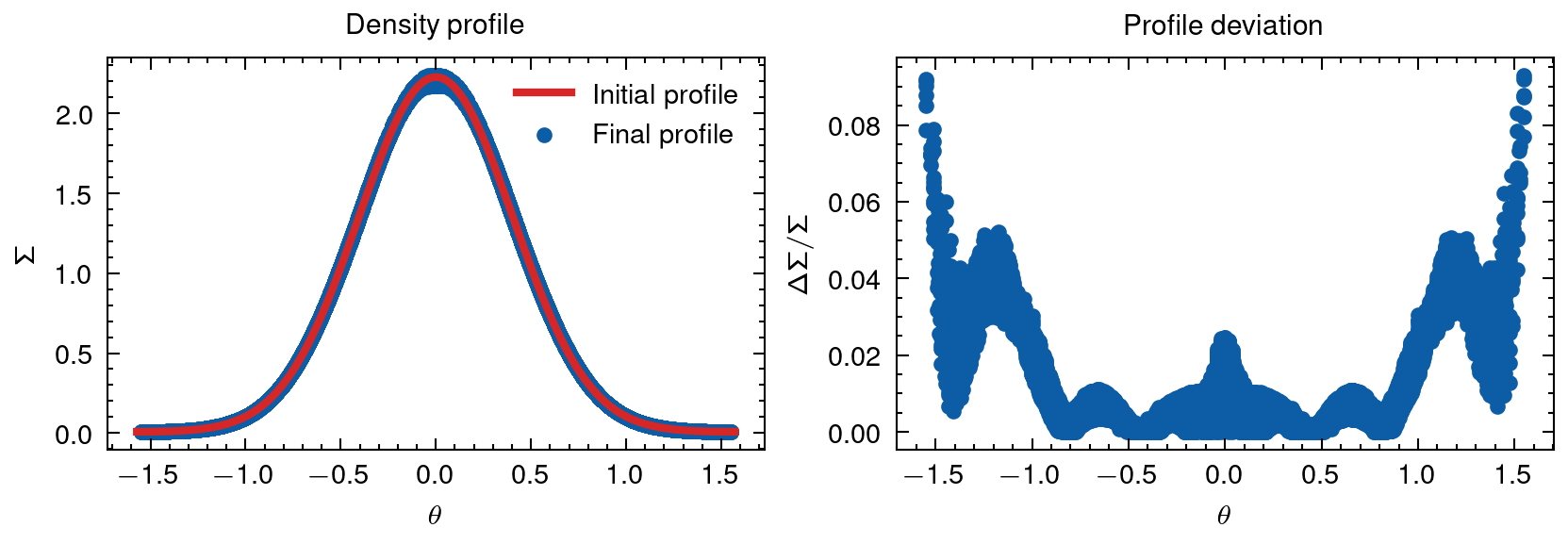}
\includegraphics[width=0.7\textwidth]{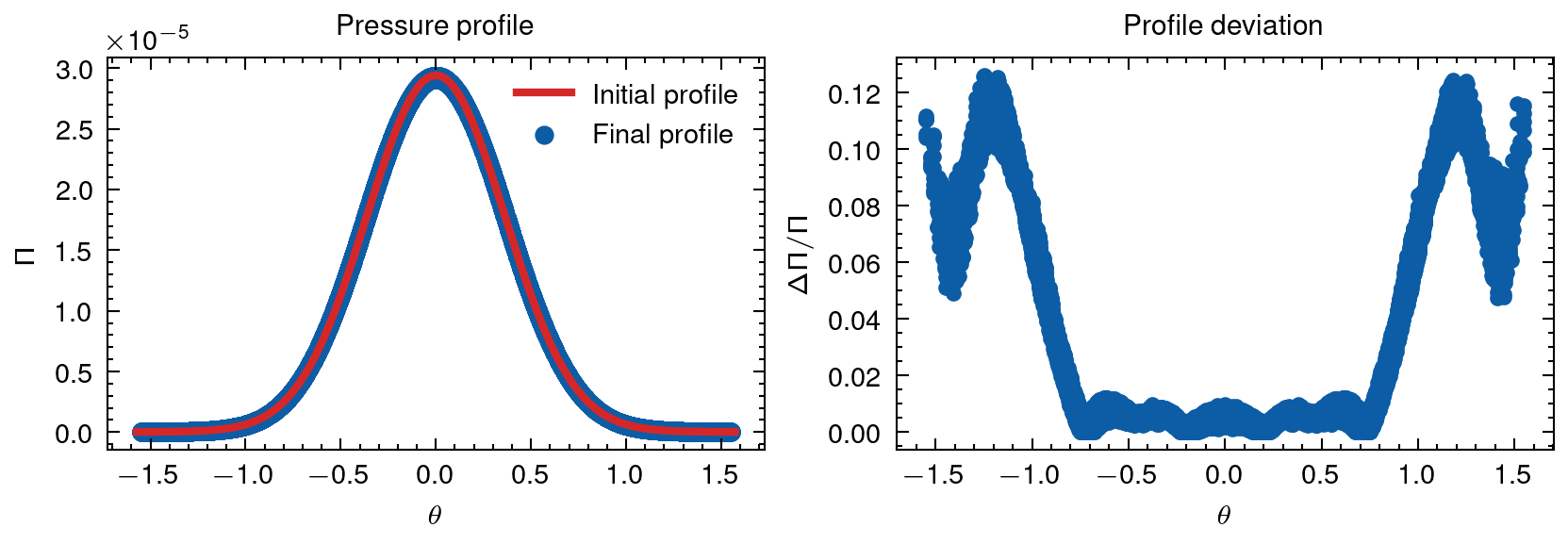}
\end{center}
\caption{Conservation of $\Sigma$ and $\Pi$ profiles after 10 full rotations on a triangular mesh with 20480 faces.}\label{fig:sigma_pi_prof_ico}
\end{figure*}

\begin{figure*} %[h]
\begin{center}
\includegraphics[width=0.7\textwidth]{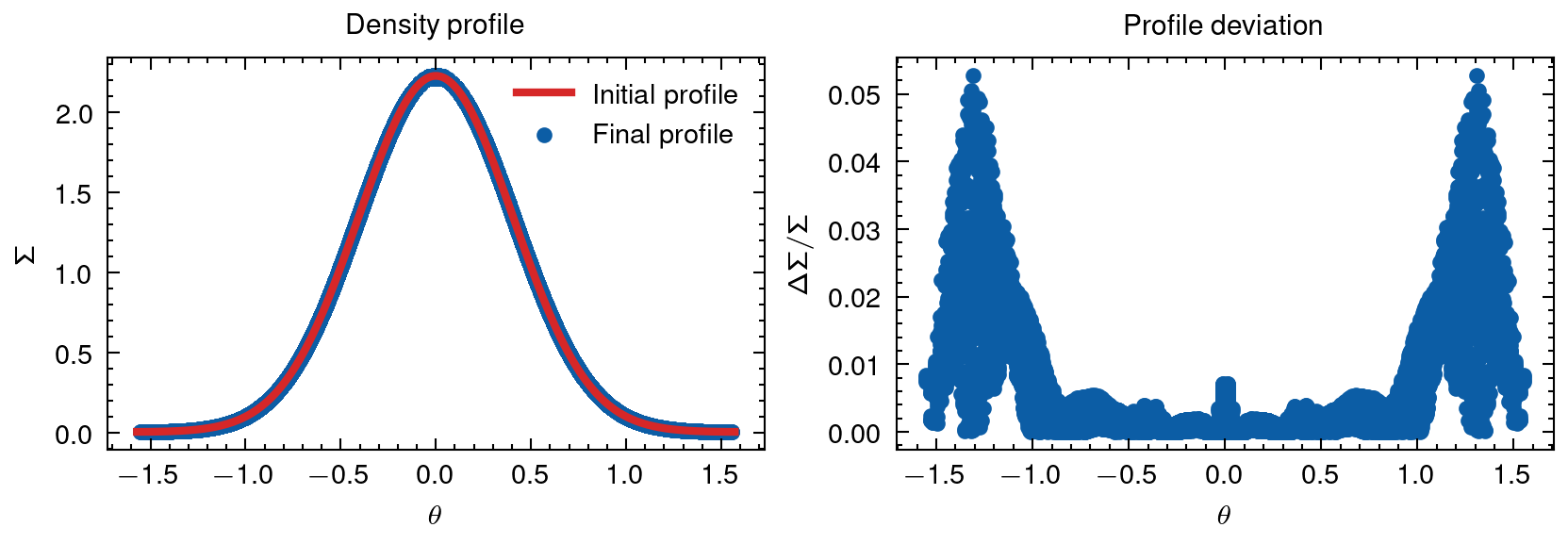}
\includegraphics[width=0.7\textwidth]{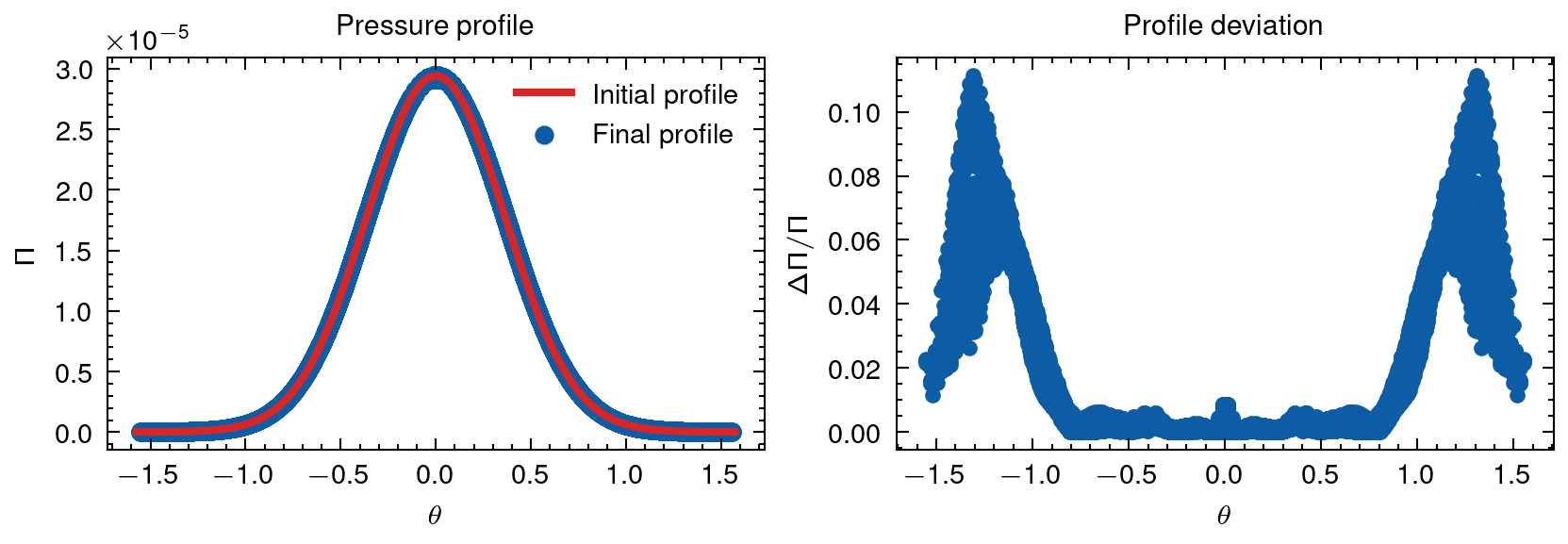}
\end{center}
\caption{Conservation of $\Sigma$ and $\Pi$ profiles after 10 full rotations on a cubic mesh with 24576 faces.}\label{fig:sigma_pi_quad}
\end{figure*}

We present the results of isentropic solution tests performed on a hexagonal, triangular, and cubic meshes with comparable number of faces. The initial pressure and density profiles have been selected as described in the isentropic solution described in Appendix~\ref{app:A}. 
The rotation is rigid-body with $\Mach_0=5$. Full simulation time is 10 full rotations.
The results show conservation of full mass and energy to the accuracy $\sim 10^{-14}$, proving that there are no numerical errors that lead to leaks or sources of conservative values (see Figure~\ref{fig:err_const_entr} for results on a  hexagonal mesh with 10242 faces). 
We expect that a physically stable steady-state solution, if set as the initial conditions for the numerical code, to remain unaltered with time.  As shown in Figures~\ref{fig:sigma_pi_prof_hex}, \ref{fig:sigma_pi_prof_ico}, and \ref{fig:sigma_pi_quad}, the profile experiences only minor deviations after 10 rotations. Here we compare hexagonal, triangular, and cubic meshes with 10242, 20480, and 24576 faces, respectively. Due to the subdivision algorithm, we cannot set arbitrary numbers of faces. With this in mind, we can only compare meshes with relatively similar numbers of faces.

The results show that hexagonal mesh yields a bigger error near the 45th parallels, yet the general profile is preserved quite well. In further tests that required isentropic configuration, we used triangular or cubic meshes.
These results allowed us to use this static profile as a baseline for further tests where we can safely add sink or source terms.
The Mach number in this solution is $\lesssim 2.5$, as rigid-body rotating isentropic configuration always retains moderate Mach numbers (see Appendix~\ref{app:A}). 

We have performed similar tests for the equal-density solution; however, it was not stable due to a convective instability. Over time this instability led to equalization of entropy, thus creating previously discussed  isentropic solution. 

\subsection{Shock test}

\begin{figure*}
\begin{center}
\includegraphics[width=0.49\textwidth]{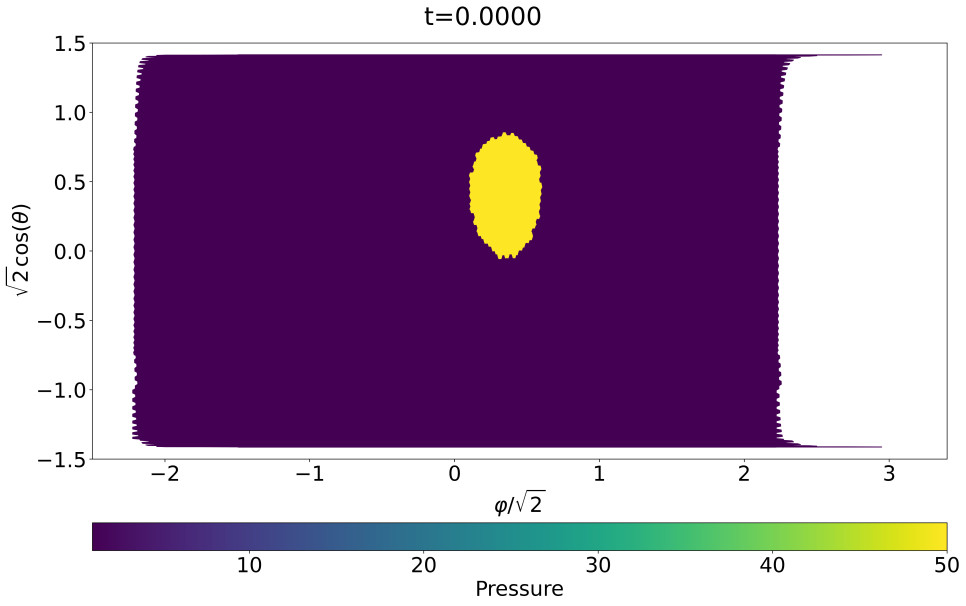}
\includegraphics[width=0.49\textwidth]{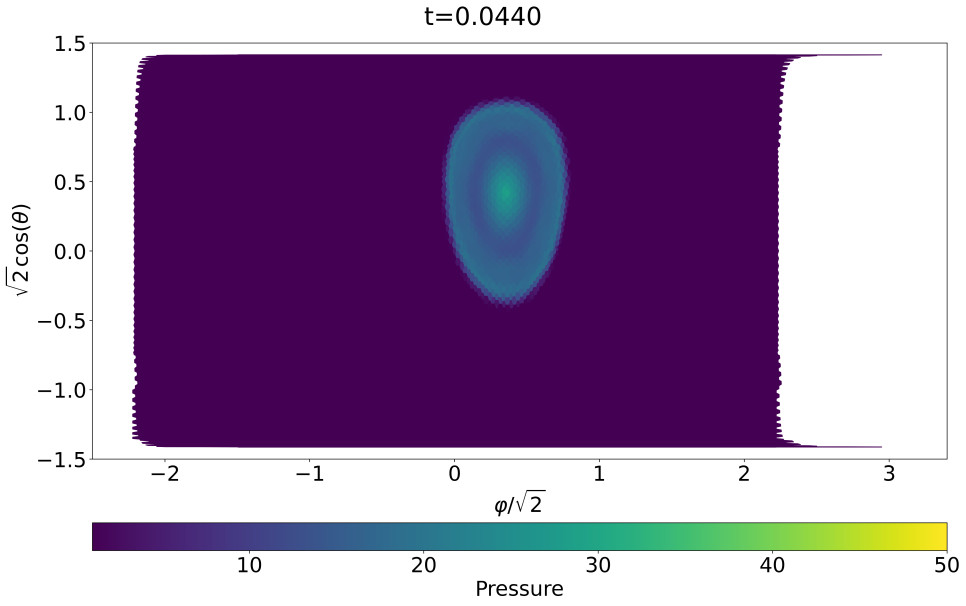}

\vspace{3mm}

\includegraphics[width=0.49\textwidth]{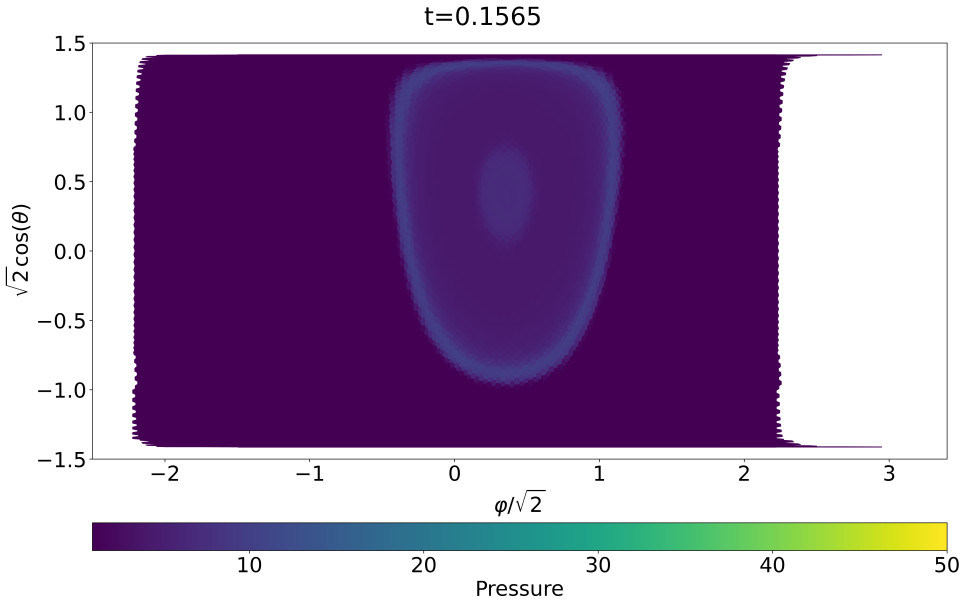}
\includegraphics[width=0.49\textwidth]{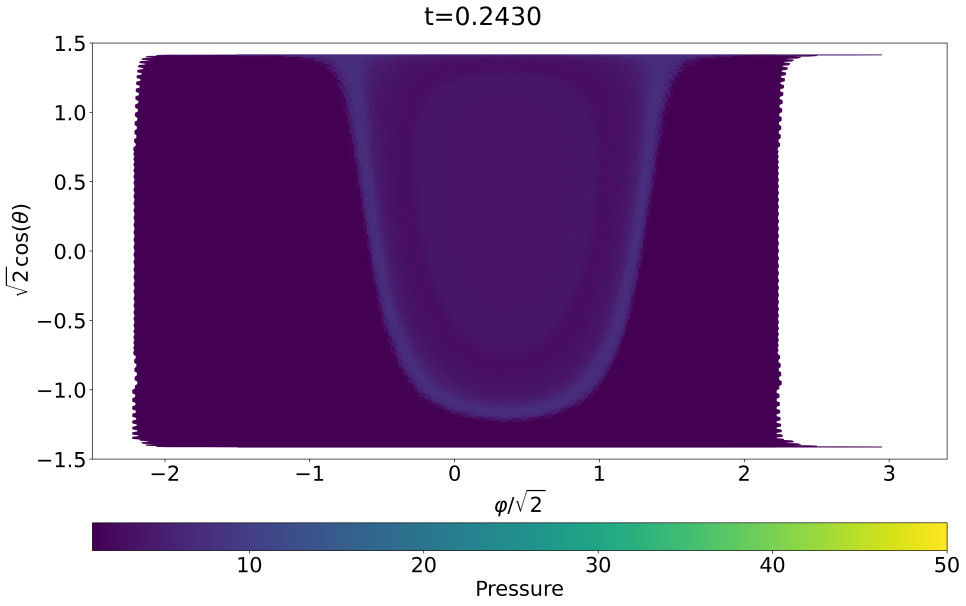}
\end{center}
\caption{Pressure maps in the shock test.}\label{fig:shock_test}
\end{figure*}

A simple shock test that has been conducted. 
The initial conditions in the test consist of a static constant-density, constant-pressure configuration with a region of the size of $R/3$ overpressured by a factor of 50. 
In order to avoid symmetry effects, the high-pressure zone was created at non-zero latitude of 17.2\textdegree \ and longitude of 28.6\textdegree. The mach number of this shock is $\Mach \approx 6$, varying depending on the time.
The results demonstrated in Figure \ref{fig:shock_test} show how the spot explodes and how the front of the shock moves over time.

Over time, the shock just bounces back and forward from the start point to the opposite side of a sphere. As we do not introduce any friction yet, it persists for some time. We continued the simulation for 3 total `bounces' and observed the initial spot compressing even further into a smaller area on a sphere before exploding once again.

\subsection{Split-sphere test}

% This test aimed at capturing the formation of vortices and the event of an inverse cascade of these vortices. Such events should turn multiple smaller vortices into a significantly larger one.

\begin{figure*}
\begin{center}
\includegraphics[width=0.49\textwidth]{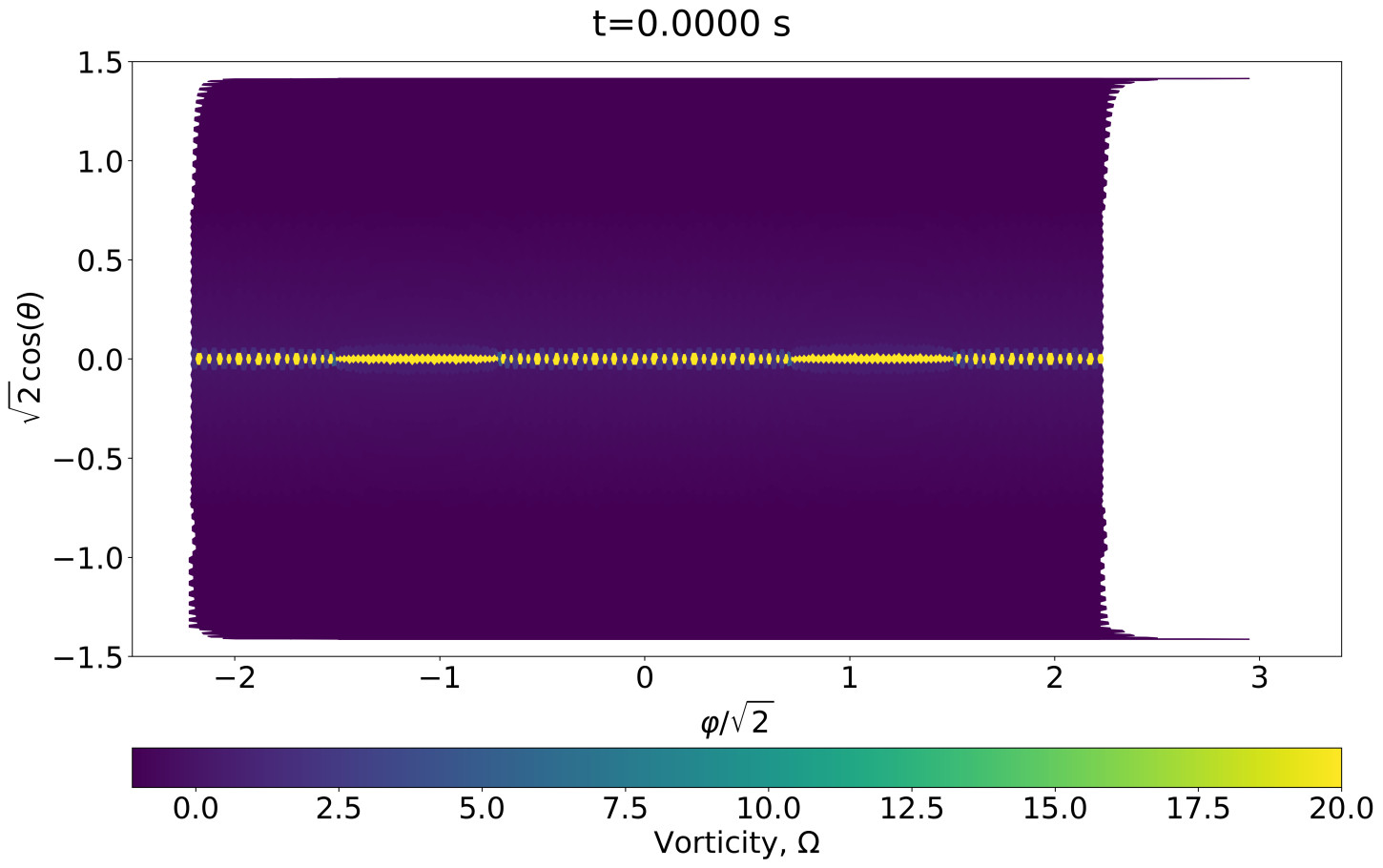}
\includegraphics[width=0.49\textwidth]{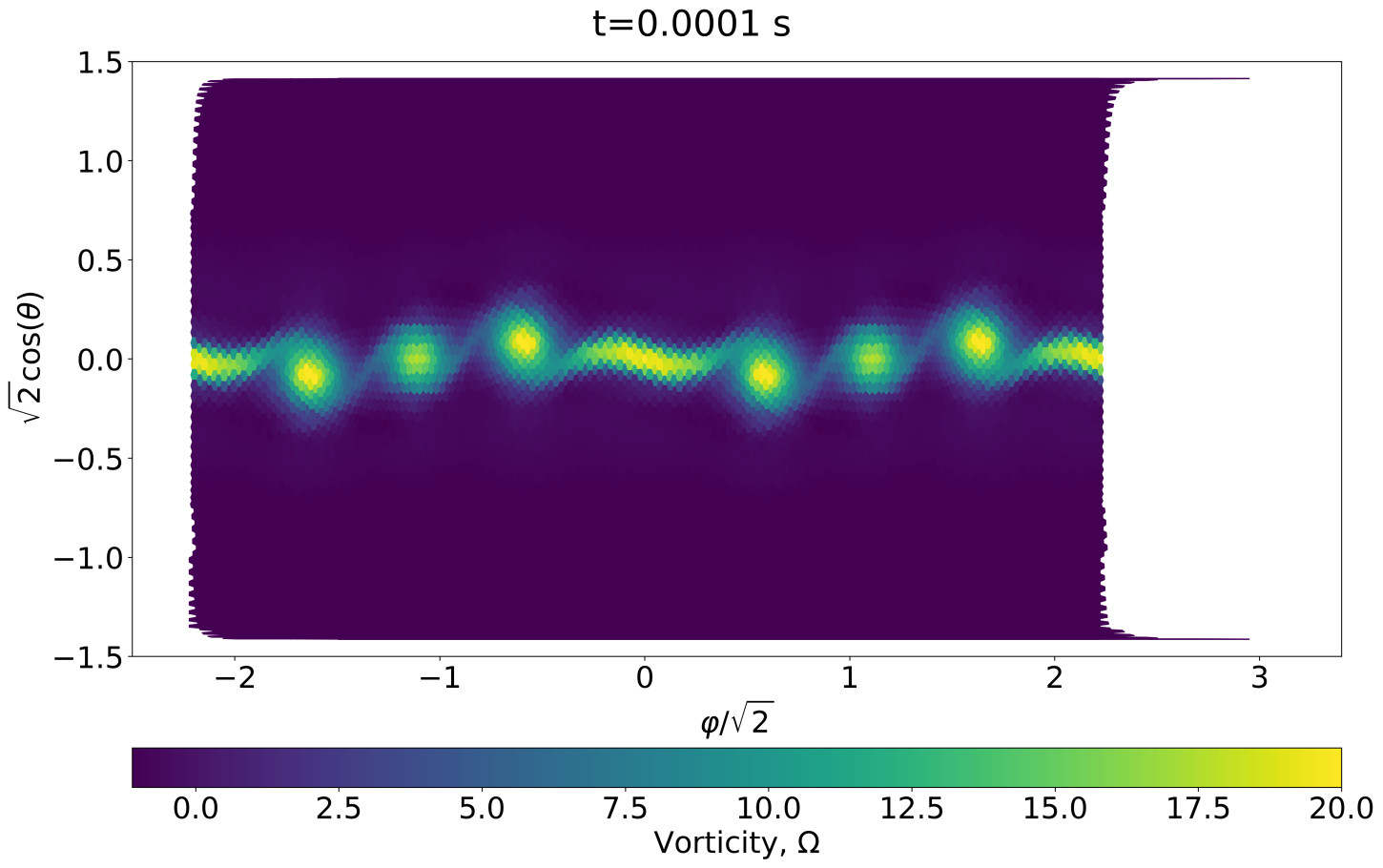}

\vspace{3mm}

\includegraphics[width=0.49\textwidth]{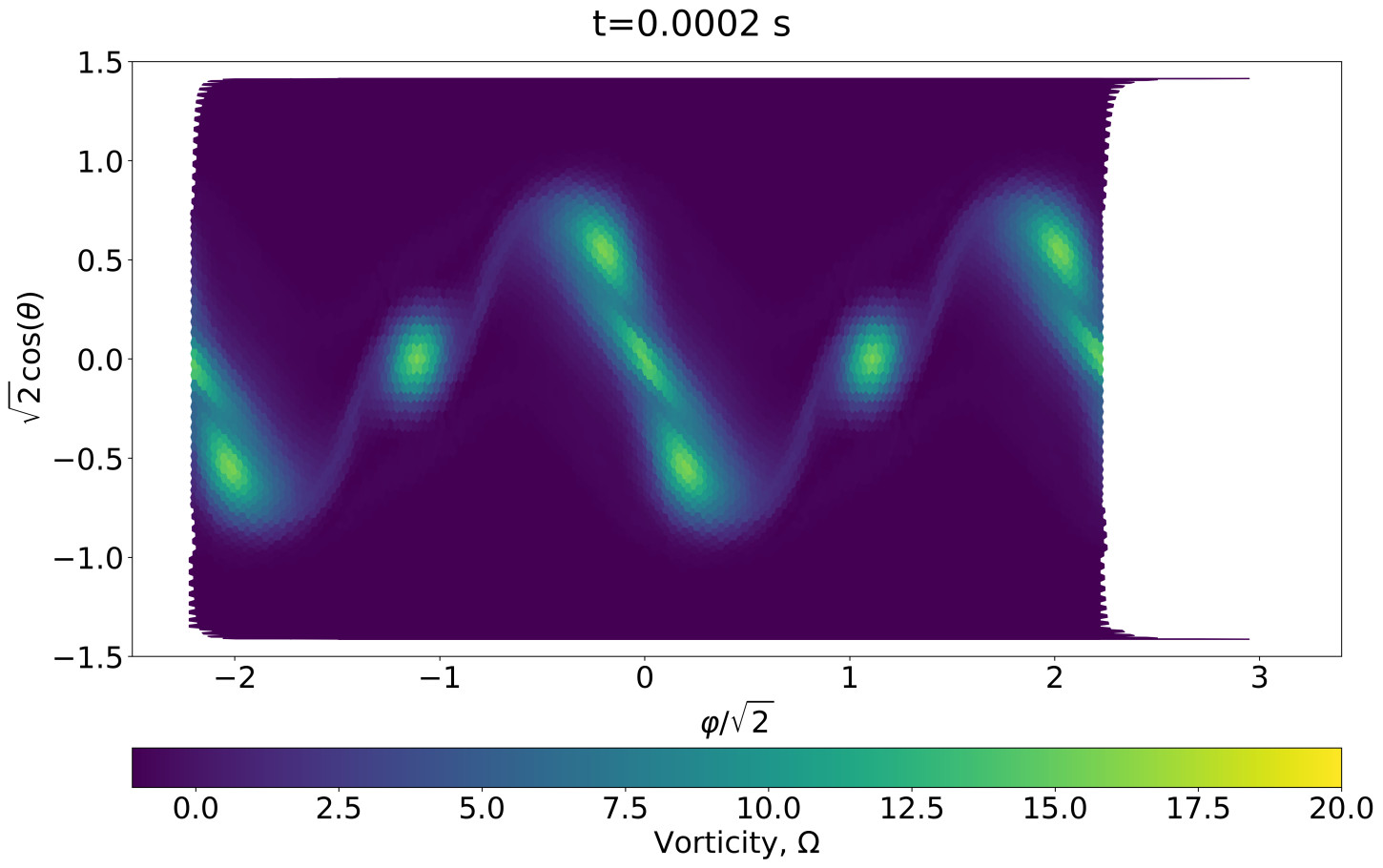}
\includegraphics[width=0.49\textwidth]{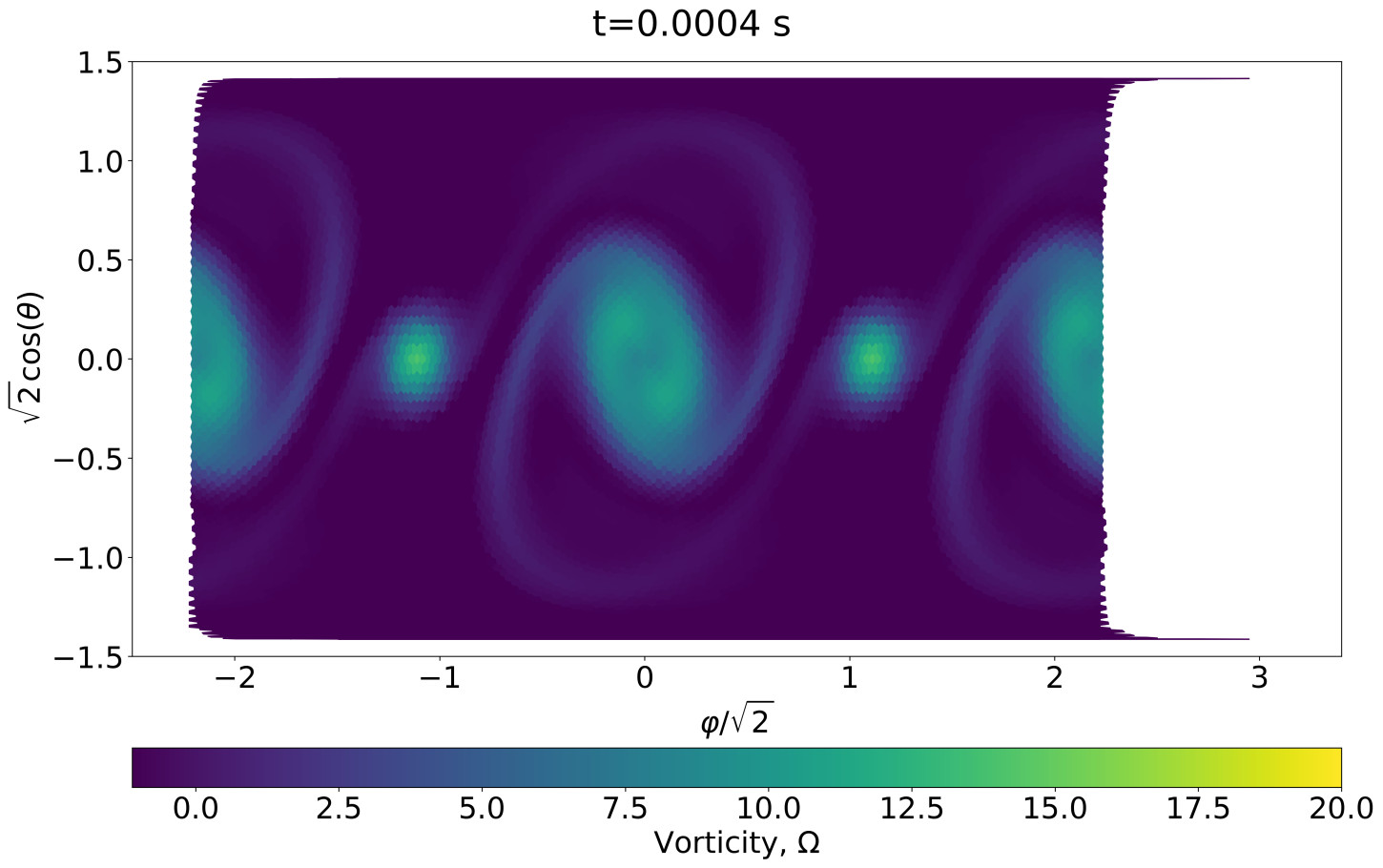}
\end{center}
\caption{Vorticity plots for the split-sphere test. The merge of multiple smaller vortices into a bigger one can be observed. }\label{fig:vort}
\end{figure*}

The initial conditions for this test consist of an atmosphere that rotates in one direction above the equator and in the opposite direction below it.
The cells on the equator are assumed to be initially static. 
Such a setup ensures that the transition is as sharp as possible, and the grid itself introduces perturbations to the axisymmetric velocity field. 
%That means faces with centers on equator are set to be static.  No smooth transition is introduced, the initial data has a discontinuity in angular momentum.
In each of the hemispheres, the rotation frequency is constant.
The discontinuity in the rotation velocity triggers Kelvin–Helmholtz instability near the equator and forms multiple small vortices that merge over time. 
%, as demonstrated in Figure~\ref{fig:vort}. There the radial component of the vorticity vector is plotted.
In Figure~\ref{fig:vort}, we show the maps of the radial component of vorticity calculated as
\begin{equation}
    \vector{\omega}_r = [\nabla \times \vector{v}]_r.
\end{equation}
This quantity is sensitive to rotations and shears in the velocity field and has the same units as $\Omega$, that allows to use the initial rotation frequency as normalization. 
Individual vortices are seen in the maps as local vorticity maxima. 
% This value allows us to observe and quantify rotational movements on the sphere. It has a natural unit of measurement: angular frequency $\Omega$.

The observed evolution towards larger spatial scales is typical for two-dimensional incompressible turbulence on a sphere. 
In particular, similar behavior was predicted in \citet{2024ApJ...971...37N} for a more complex setup with several streams rotating in alternating directions. 
% The authors predict merges and growth in the size of the vortices. % Something similar can be seen in our results, although our initial conditions for this problem are different. 
% Nonetheless, these results show that our code is capable of modeling this phenomenon, and this opens up a possibility for further study.

\section{Accretion spreading layer formation}\label{sec:SL}

\subsection{Simulation setup}\label{sec:SL:num}

Here we combine the initial equal-entropy atmosphere described in the Section~\ref{sec:res:stat} with additional source terms representing accretion and other physical effects from Section~\ref{sec:phys:src}. Additionally, we perform this computation with variable $\gamma$ to model mixed-gas structure of a spreading layer.
The initial state is an equal-entropy atmosphere (see Appendix~\ref{app:A}) with the density at the poles $\Sigma_0 = 7.7 \times 10^5 \unit{\gram\per\square\centi\metre}$, $\Mach_0 = 5$, and rotation frequency $\Omega = 300 {\rm s}^{-1}$. 
Equatorial rotation velocity is thus $\sim 0.01c$.
The average surface density for this solution, assuming $\Gamma_{\rm start} = 7/5$,  is $\langle\Sigma\rangle =3\cdot 10^6 \unit{\gram\per\square\centi\metre}$. 
% The Mach number of the initial state is $\Mach_0 = 5$ with equatorial velocity being equal to $0.01 \rm c$. 
We also assume the speed of sound on the equator to be equal to $2 \times 10^{-3} c$. We do not incorporate any NS deformation into our computation for now, so $\Omega_0=0$.

We can assume equal-entropy atmosphere, as the accretion rate in LMXBs is highly variable, and the case of a reduced accretion rate leaving a neutron star by itself to form a steady state atmosphere. Rotation frequency was selected relatively low to improve code stability.

% For the accretion, w
We set the accretion rate to a $10^{-8} M_{\odot} \ \rm yr^{-1}$, with the equatorial velocity $|\vector{v}_{\rm orb}| = 0.4 c$, which is close to the Kepler velocity of $0.47 c$. $ \left( \frac{E}{\Sigma} \right)_{\rm d}$ is set to $6 \times 10^{-3}$ in code units.
Accreted mass is distributed over the sphere according to the Gaussian law (equation~\ref{E:source:Gauss}), where we set $\alpha_{\rm tilt} = 6^\circ$ and $\sigma = 6^\circ$. 
Tilting the source allows to break the axial symmetry of the problem. From the physical point of view, this distribution corresponds to a thin disk with the thickness $\sim 0.1R$. 
Expected standard disk thickness at about the last stable orbit is \citep[equation 2.8]{1973A&A....24..337S}
\begin{equation}
    \left(\frac{H}{R} \right)_{\rm st} = \frac{3}{8\uppi} \frac{\varkappa \dot{M}}{c R} \left( 1-\sqrt{\frac{R_{\rm in}}{R}}\right),
\end{equation}
that varies from $\sim 0.5$ away from the inner radius $R_{\rm in} \simeq 6GM/c^2$ to essentially zero near this radius. Making more accurate estimates of the disk thickness in this region would require a more accurate model of its transonic part near the last stable orbit. 
% We add a 6\textdegree \ tilt to the accretion disk with respect to the spreading layer's equator. The accretion has a normal distribution on the surface with a $\rm FWHM = 0.1 R$, resulting in a geometrically thin accretion disс.

We used a cubic mesh with 24576 faces. The simulation covers a time period of $ \sim 1\unit{\second}$. 
% It is not enough to model the spreading layer because in order for the initial mass of the envelope to be replaced by accreted material we would need to run simulation for $ \sim 100 \unit{\second}$.
A realistic steady-state model of a SL requires a much longer simulation with a duration at least several mass renewal time scales ($\gtrsim 100\unit{s}$). 
A longer simulation would ensure mass equilibrium. 
Besides, the code currently does not have any friction with the NS surface, that is a crucial constituent for angular momentum conservation and energy release. 
We are planning to present a more realistic model in a separate paper. 
Here, we rather expect the simulation to reflect the dynamic-timescale phenomena such as oscillations and instabilities that are expected to emerge during the formation of an SL or a rapid increase in mass accretion rate. 
% Thus, the results presented can only describe the initial formation of the spreading layer.
Results of the simulations are stored as snapshots of all the primitives in all the cells saved several times per rotation period and as high-cadence `light curves' representing the properties of the object from the point of view of a distant observer (see later Section~\ref{sec:SL:timing}). 
% The main results of this simulation are stored in the modeled light curves for observers with different geometries. Three types of observer were computed, one on the axis of rotation, one perpendicular to it, and one in between. From this light curves a dynamic power spectrum has been computed, showing the way different modes have evolved over time. 

\subsection{Overall dynamics of SL development}\label{sec:SL:dyn}

\begin{figure*}[h]
\begin{center}
\includegraphics[width=0.49\textwidth]{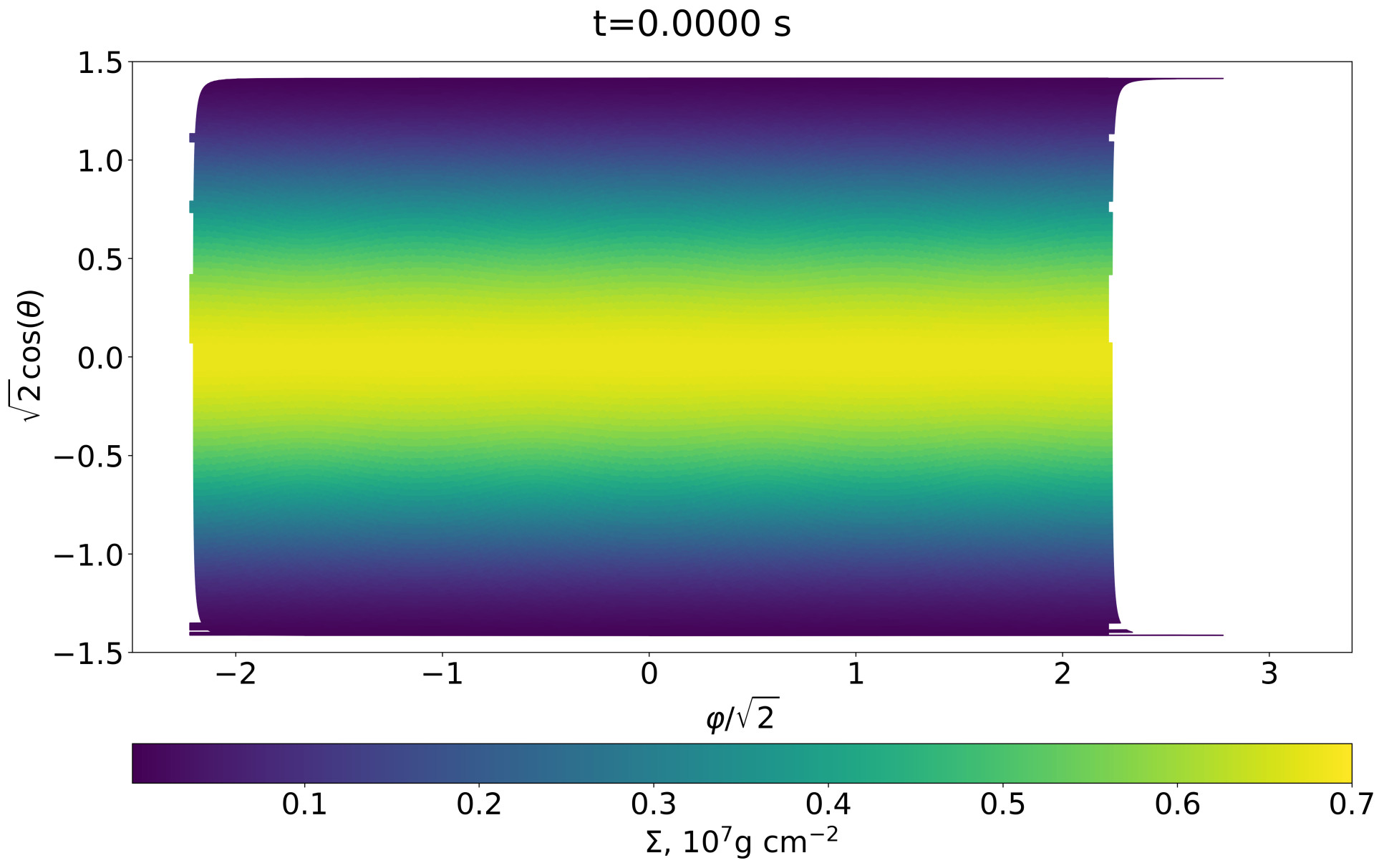}
\includegraphics[width=0.49\textwidth]{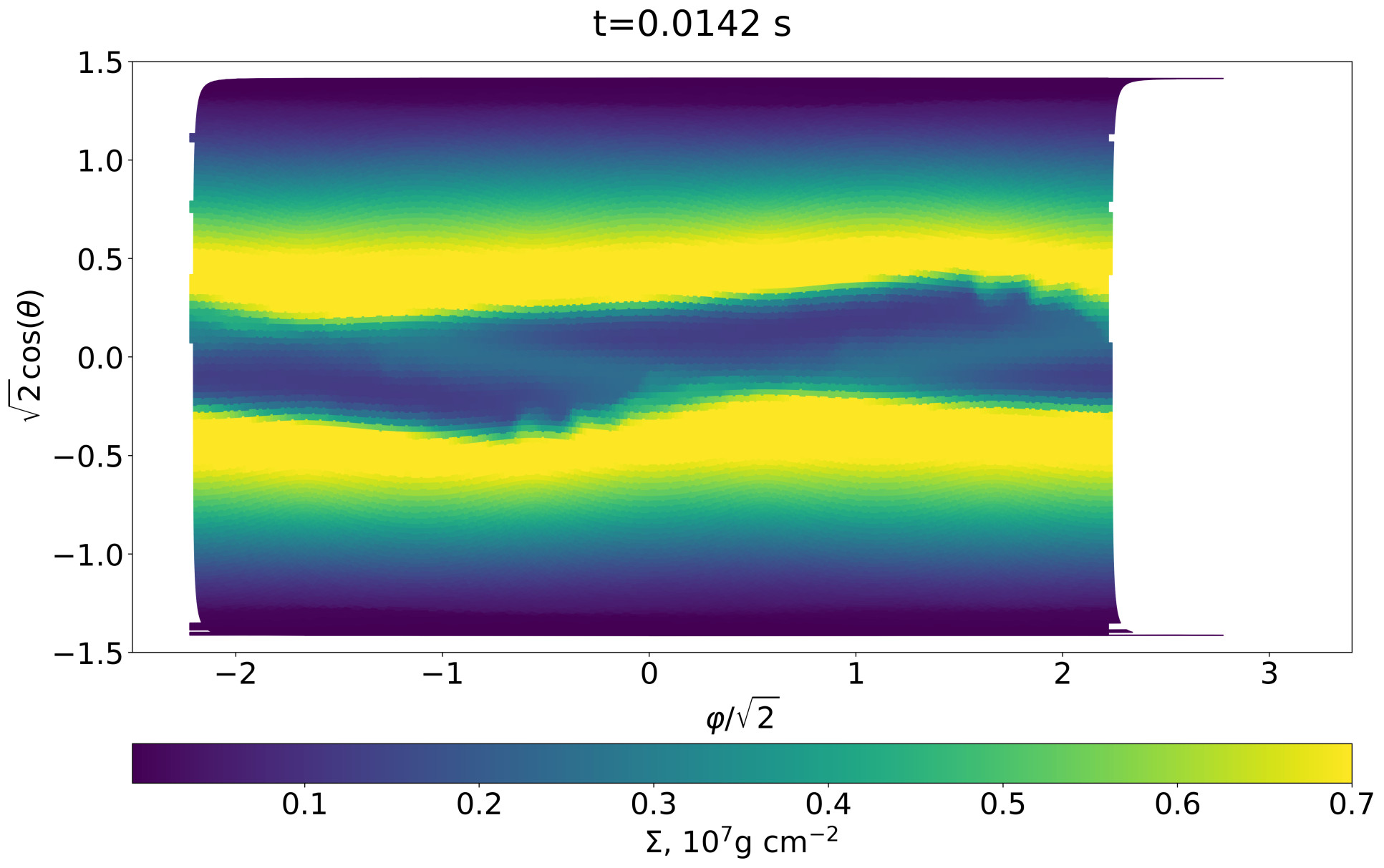}

\vspace{3mm}

\includegraphics[width=0.49\textwidth]{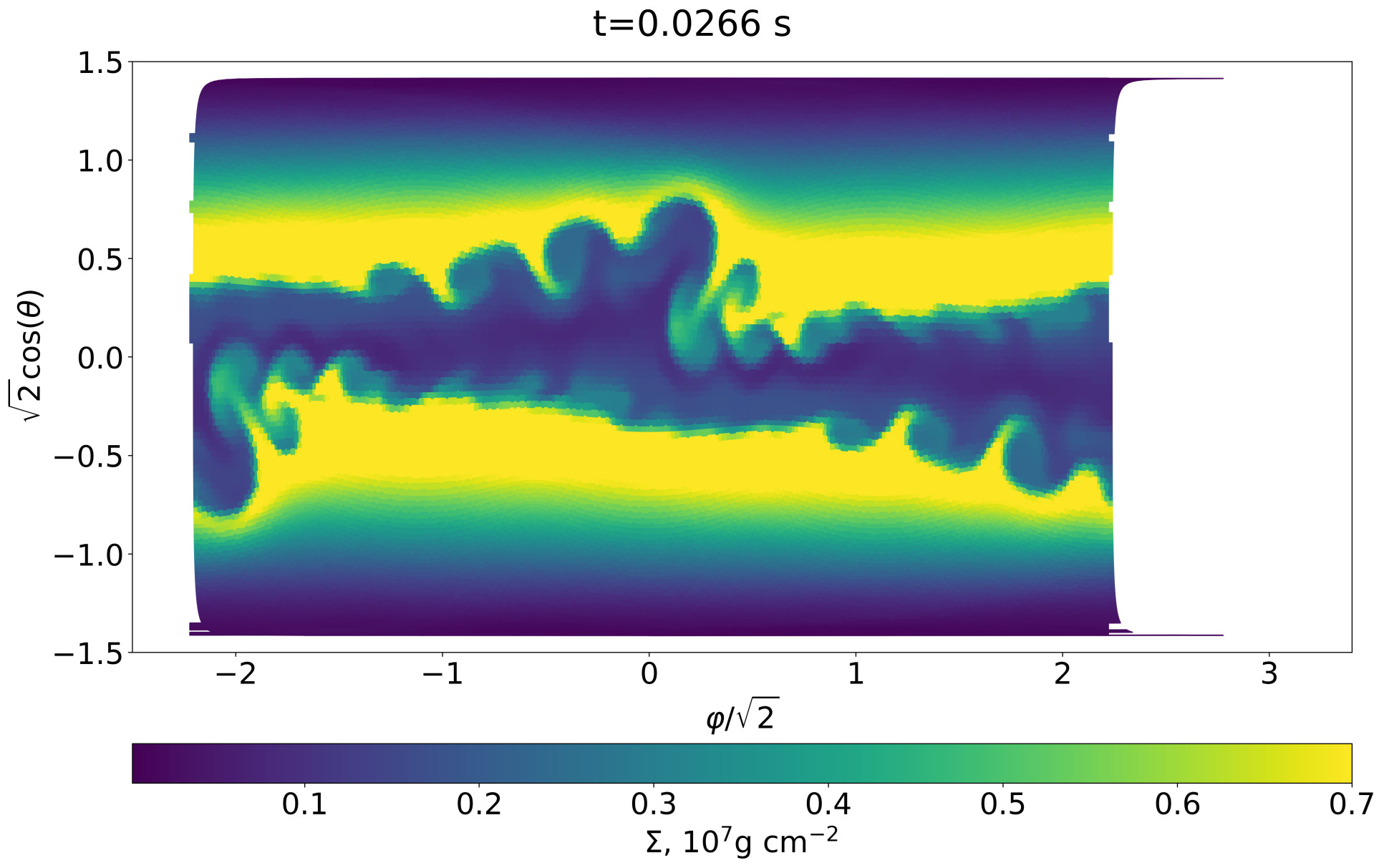}
\includegraphics[width=0.49\textwidth]{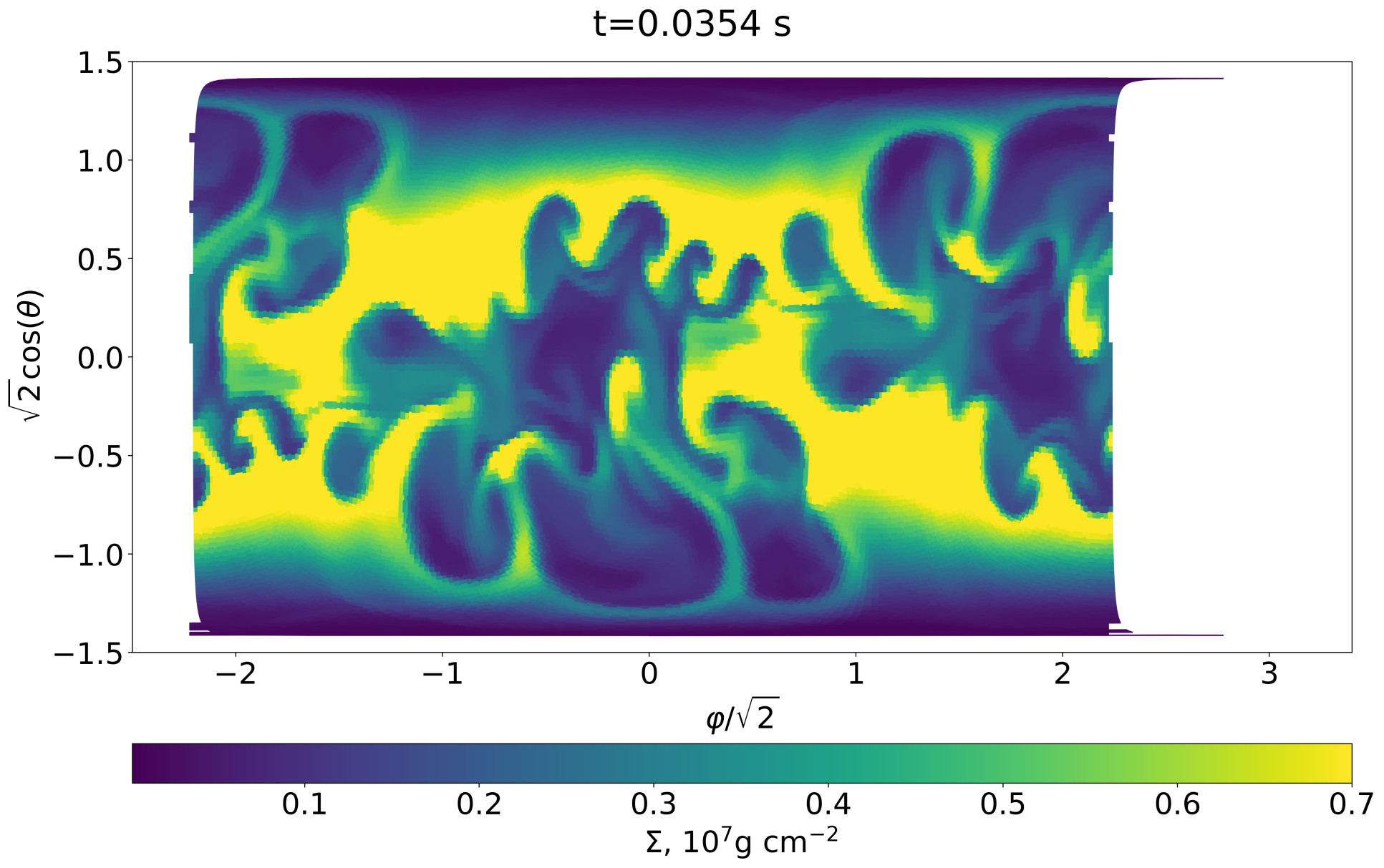}

\vspace{3mm}

\includegraphics[width=0.49\textwidth]{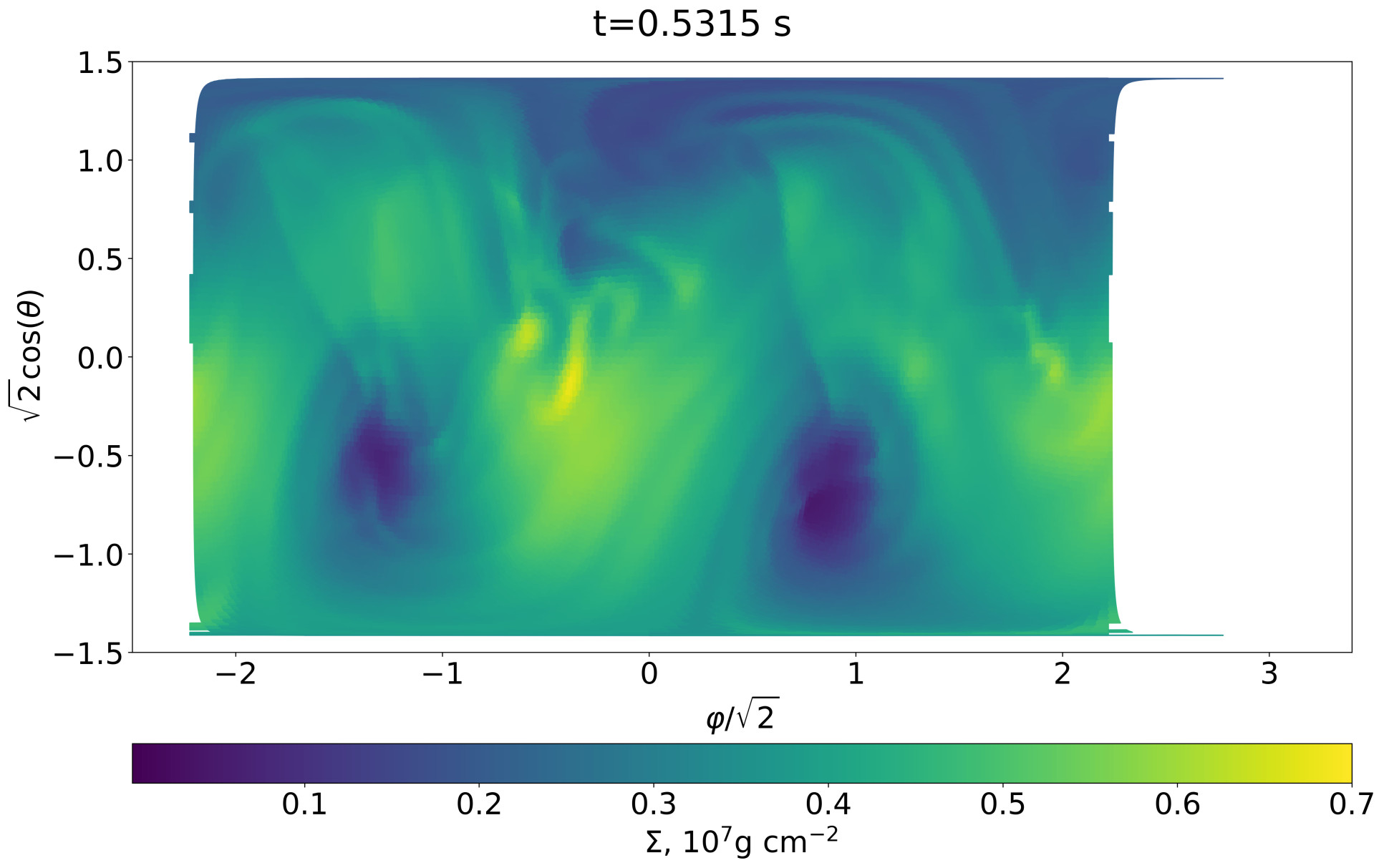}
\includegraphics[width=0.49\textwidth]{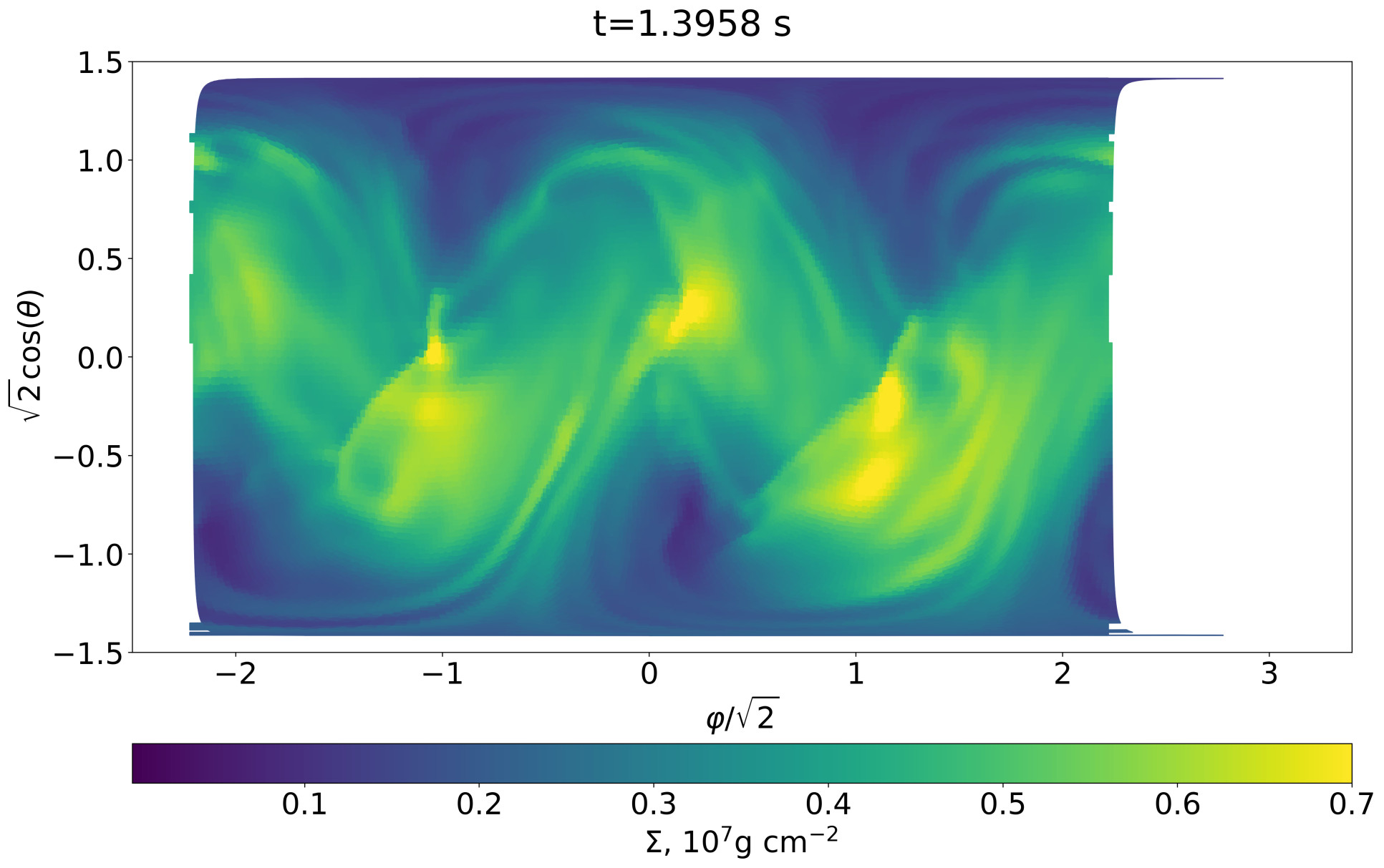}
\end{center}
\caption{Density maps of the accretion simulation. Six snapshots: initial state, instability start, instability evolution, and the final state. Full video of simulation is available on YouTube: \url{https://youtu.be/BVzGHAnDtwA}.}\label{fig:big_sim}
\end{figure*}

\begin{figure*}[h]
\begin{center}
\includegraphics[width=0.99\textwidth]{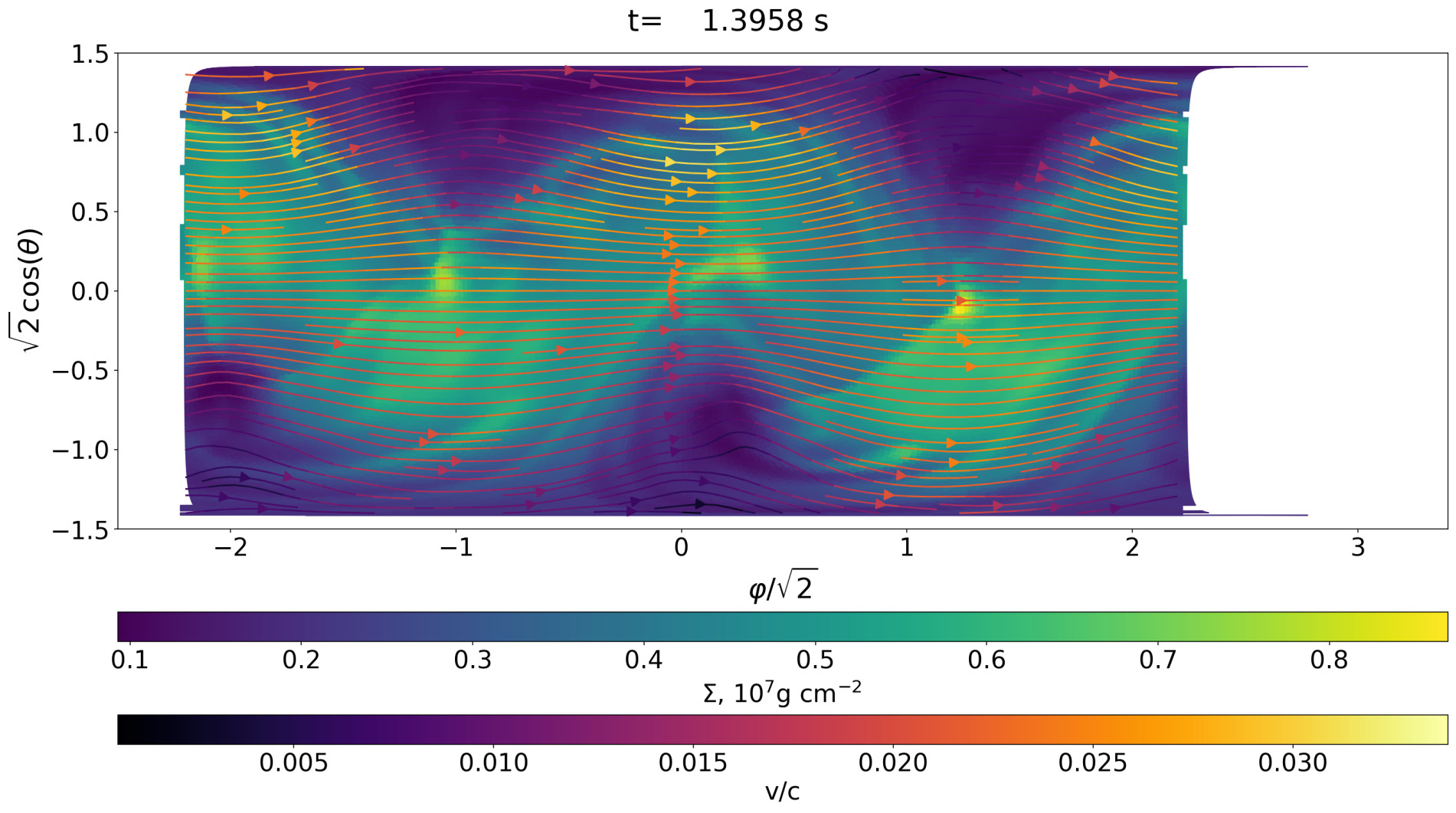}
\includegraphics[width=0.99\textwidth]{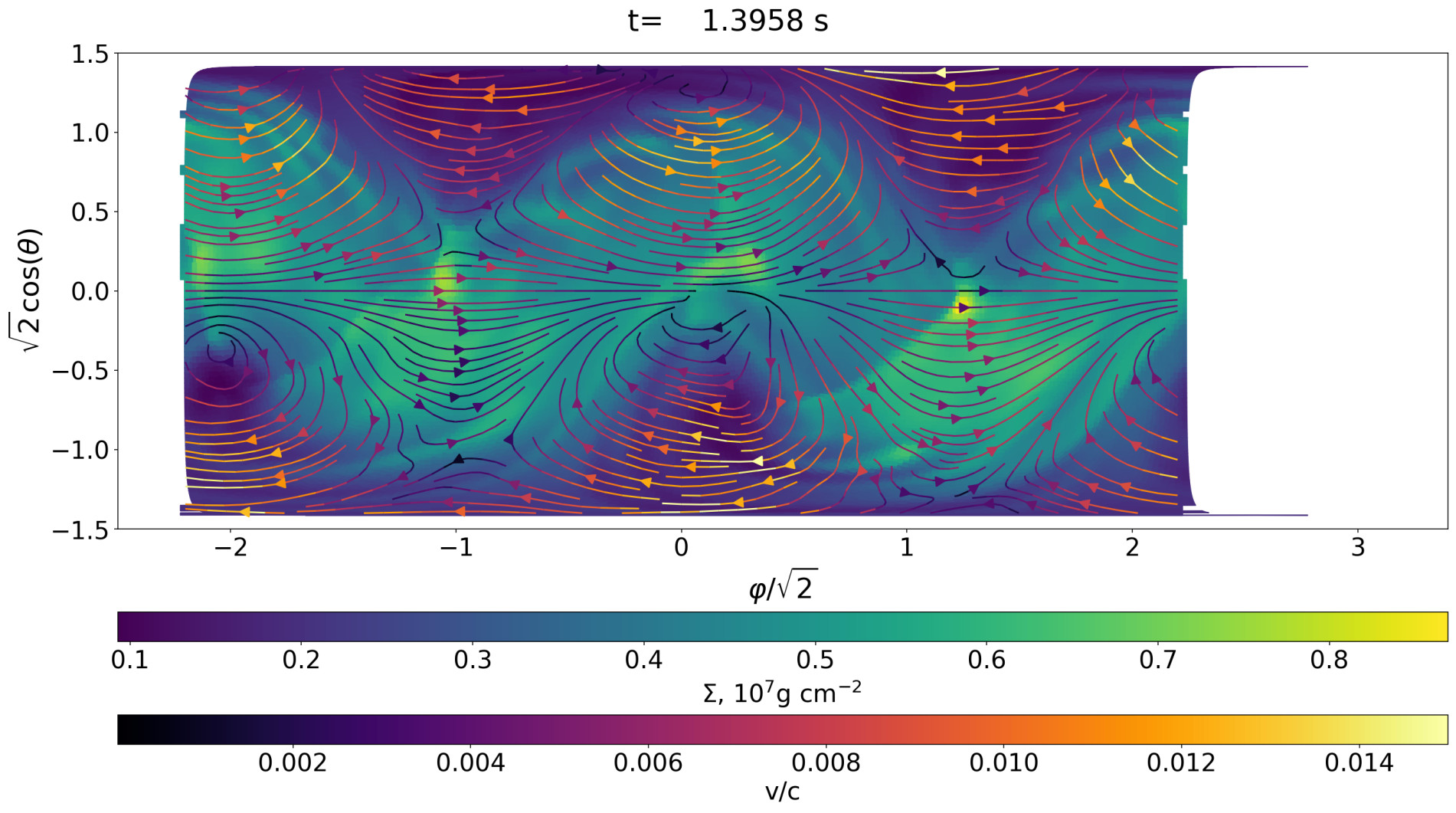}
\end{center}
\caption{Final state of accretion simulation with streamlines. First plot contains streamlines with rotation, the second is without average rotation. }\label{fig:stream}
\end{figure*}

The simulation passes several stages shown in Figure~\ref{fig:big_sim}. 
Accretion breaks the initial state by heating the equatorial regions and creates conditions for convective instability.
The latter develops on the time scales of $\sim 1/\Omega$ and by $t\sim 0.05\rm s$ enters a nonlinear stage. 
As we expect for a Rayleigh-Taylor instability, the small-scale modes develop faster but are overrun by longer-wavelength modes, that finally merge to form the global `tennis ball' pattern (at around t$\sim 0.2$s) clearly visible in the last two panels of Figure~\ref{fig:big_sim}. 
The pattern rotates around the $z$ axis at a frequency close to the average angular frequency. 
During the later stages of evolution, this pattern is surprisingly stable and rotates at a frequency close to the initial rotation frequency of the layer. 
The low-density, low-pressure regions located within the meanders of the pattern are long-lived cyclones with their own local velocity fields.
In Figure~\ref{fig:stream}, we show two density maps for the last snapshot of the simulation together with the velocity streamlines. 
In the second panel, the average rotation is subtracted from the velocity field, that makes the velocity circulation patterns in the cyclones clearly visible. 
% with rotation and without it.
In order to subtract rotation for the second picture in Figure~\ref{fig:stream}, we compute average linear rotation frequency as
\begin{equation}
\nu = \frac{1}{2\uppi N} \displaystyle \sum_i (\vector{R}_i \times \vector{v}_i)_z \csc^2 \theta_i,
\end{equation}
where $N$ is the total number of cells in the mesh.
The linear frequency can be turned into rotational frequency by multiplying it by $2\pi$. After that, we compute rotational velocities and subtract them from the overall velocity field.

This subtraction allows to resolve the circular pattern of two cyclones in the southern hemisphere, with the strongest one located at $\theta \sim -60^\circ$ and $\varphi \sim 200^\circ$. 
After $t\sim 0.35$s, the strongest cyclone coincides with the global minimum of pressure, which allows one to accurately trace its position and measure the rotation velocity of the `tennis-ball' pattern. 

\begin{figure}
\begin{center}
\includegraphics[width=0.49\textwidth]{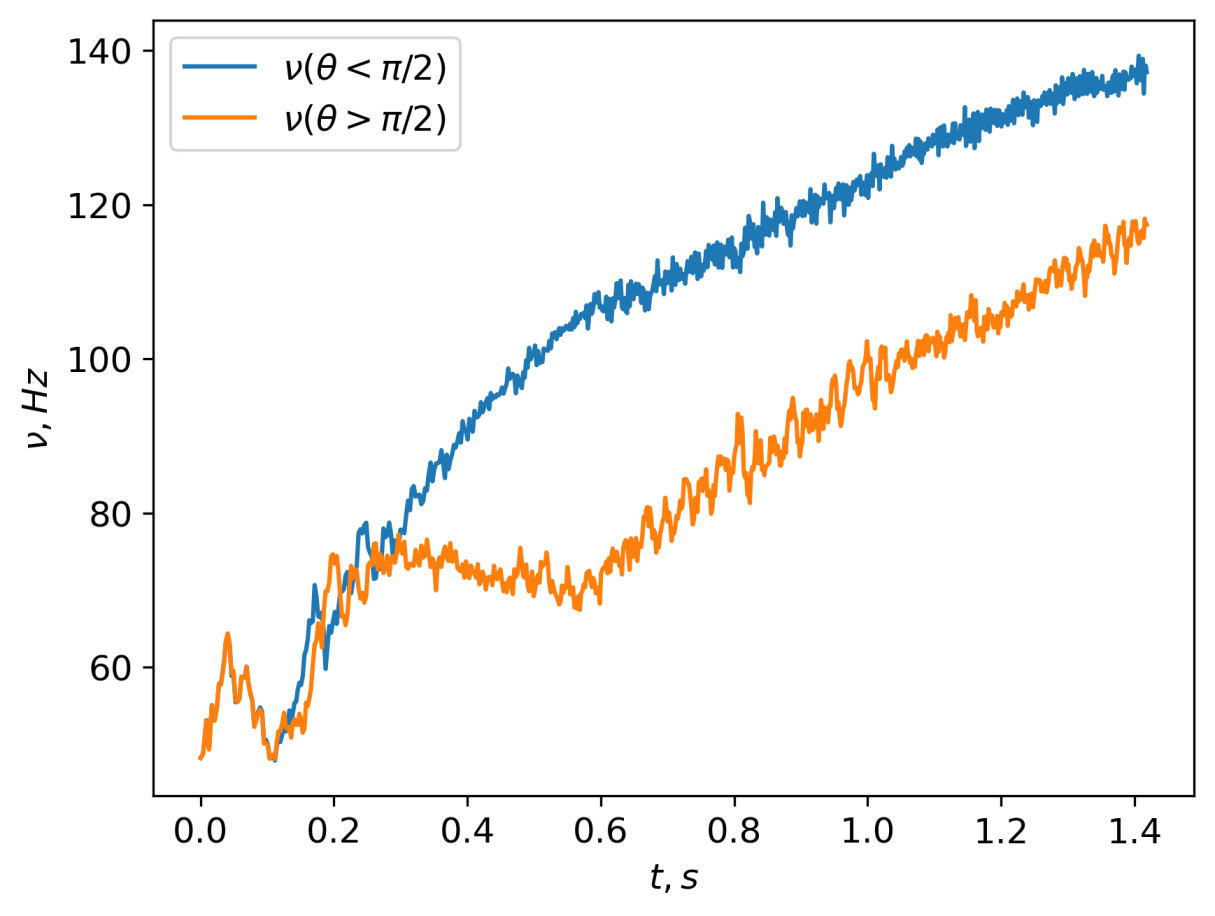}
\end{center}
\caption{Average rotational frequency of northern and southern hemispheres.}\label{fig:split}
\end{figure}

It is important to note that rotation at later stages of the simulation is not uniform and varies with latitude. 
In particular, rotation frequencies at later stages of the evolution differ considerably for the northern and southern hemispheres, as demonstrated in Figure \ref{fig:split}.
% For example, averaging out rotations in northern and southern hemispheres leads to a noticeable split in frequencies, as demonstrated in Figure \ref{fig:split}.

% last snapshot of the simulation with the velocity stream lines overlapped. 
% The pronounced minimum of surface density observed in the southern hemisphere has also an excess in vorticity, hence we can classify it as a long-living cyclone. 
% and creates instabilities (we assume them to be Rayleigh–Taylor instabilities). 
% As they evolve, they form a clear sin-like structure that can be seen in Figure \ref{fig:stream}. This structure continues rotating almost uniformly on the surface, contributing to what we see in the power spectra.

% In the southern hemisphere, we can observe a cyclone. It formed during the initial instability and then persisted over a lot of rotations up until the end of the simulation. 

\subsection{Timing analysis}\label{sec:SL:timing}

We compute light curves as follows: let $\vector R_{\rm obs}$ be the unit vector setting the direction towards the observer.
% We write the light curves for three observers: equatorial, polar, and intermediate one. 
Let $\zeta$ be a set of faces with radius vectors $ \vector R_i$, such that $\vector R_{\rm obs} \cdot \vector R_i > 0$. Then for each time step we find the `flux' value for the light curve as
\begin{equation}
    L_{\rm c}= \displaystyle \sum_{i \in \zeta} \Pi_i K_i \left( \vector{R}_{\rm obs}\cdot \vector{R}_i\right),
\end{equation}
% where $\alpha_i$ is the angle between $\vec R_{\rm obs} $ and $ \vec R_i$, 
where $K_i$ is the surface area of the $i$-th cell, and $\vector{R_i}$ is the normalized radius vector of its center given by equation~(\ref{eq:face_center}). 
%Note that this is just a pressure-based approximation for the real light curve; nevertheless, it does show major trends. We also tested light curves created using radiative losses from \ref{E:sink:rad}; however, these turned out to be too spectrally blurry, and that led to the loss of certain harmonics.
Calculating such a time series for weighed pressure allows us to outline the most dynamically violent processes in the layer, which do not necessarily affect the radiative energy loss term (equation~\ref{E:sink:rad}). 
% We leave a more elaborate study of spectral variability within our model for a separate study.
The series are calculated for four different observers with the inclination angles $i = 0$ (along the initial rotation axis, North pole), $45^\circ$, $90^\circ$ (in the equatorial plane), and $180^\circ$ (South pole). 

As a result, we have four slightly inhomogeneous time series  of $\sim 5\times 10^5$ data points each. The inhomogeneity is caused by the variable time step in equation (\ref{eq:dt}) taking effect during some parts of the simulation. This leads to the spacing between points of the time series being slightly uneven. 
% a loss of even spacing between points of the time series.
% Writing down the value of $Lc$ at every step gives us the time series that we can perform time series analysis on.

During the simulation, the spreading layer is out of thermal balance, and all the variability curves are dominated by accretion-related heating. 
As a result, all the light curves are demonstrating a strong growing trend that needs to be removed to study the variability on smaller time scales (of the order sound propagation time and rotation period).
We subtract the trends using multiple connected splines of the third order, split our observations into 10 even (in terms of the amount of data points) segments, and then compute a periodogram for each of them. 
We use the Lomb-Scargle algorithm \citep{1976Ap&SS..39..447L, 1982ApJ...263..835S} to account for the inhomogeneity of the timing data. 
This split allows us to see changes in the power density spectrum over time.

One of the most prominent features of the simulation is the presence of two cyclones in the southern hemisphere. They are part of the rigidly rotating pattern formed as the result of the convective instability. We use the larger of the cyclones to track the rotation of the pattern itself. We determine their rotation frequency $\nu_{\rm cycl}$ by finding the global pressure minimum and tracking its motion by searching for a local pressure minimum in the vicinity of the location found for the previous frame.

%As the rotation stops being rigid-body with noticeable frequency difference between hemispheres starting around $t\sim 0.1$s, as can be seen in Fig. \ref{fig:split}, it becomes more important to track the frequency of the cyclones rotation. We can assess the $\nu_{\rm cycl}$ by tracing the center of the larger of the two cyclones over multiple snapshots, as it creates a local minimum in pressure and maximum in vorticity. 

\begin{figure*}
\begin{center}
\includegraphics[width=0.49\textwidth]{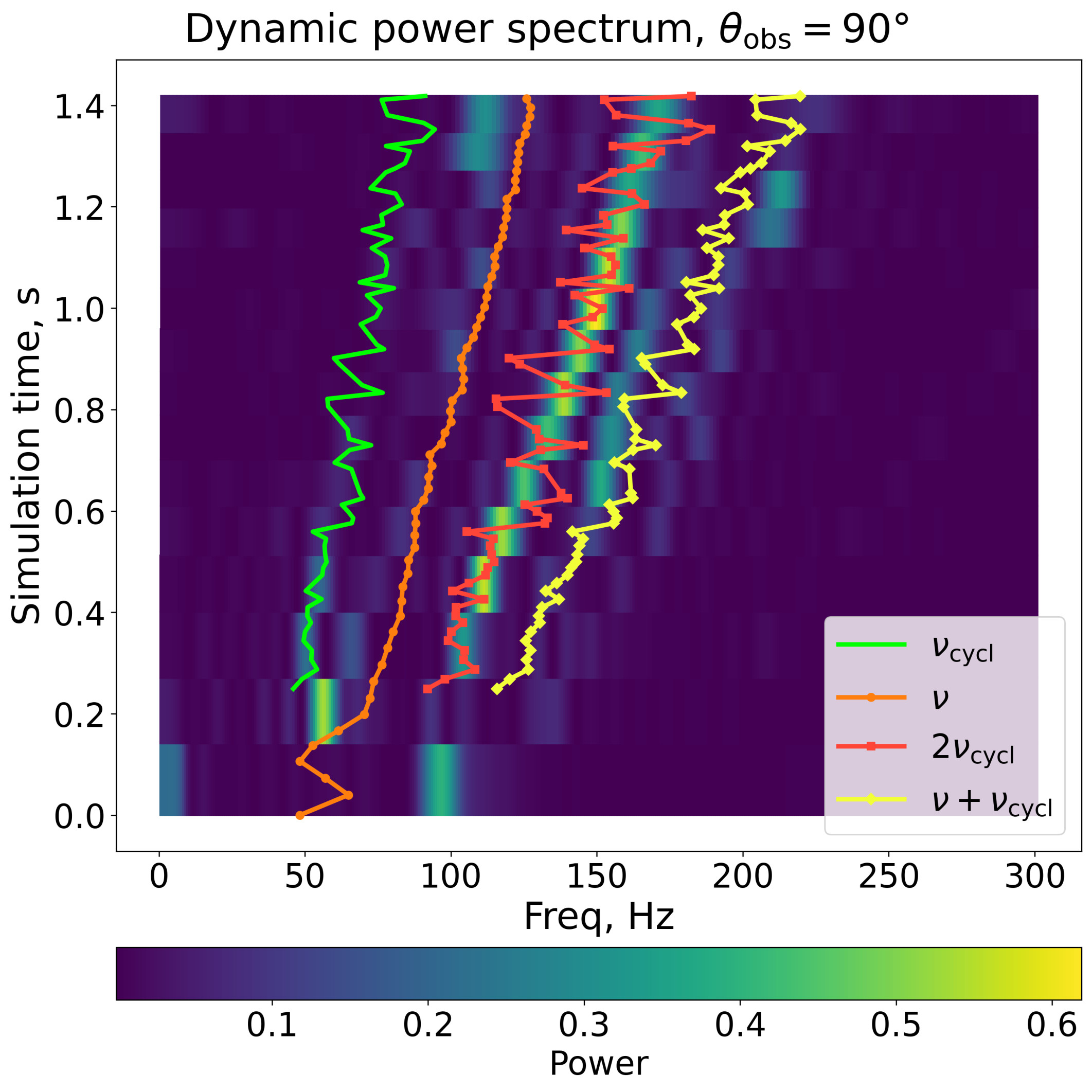}
\includegraphics[width=0.49\textwidth]{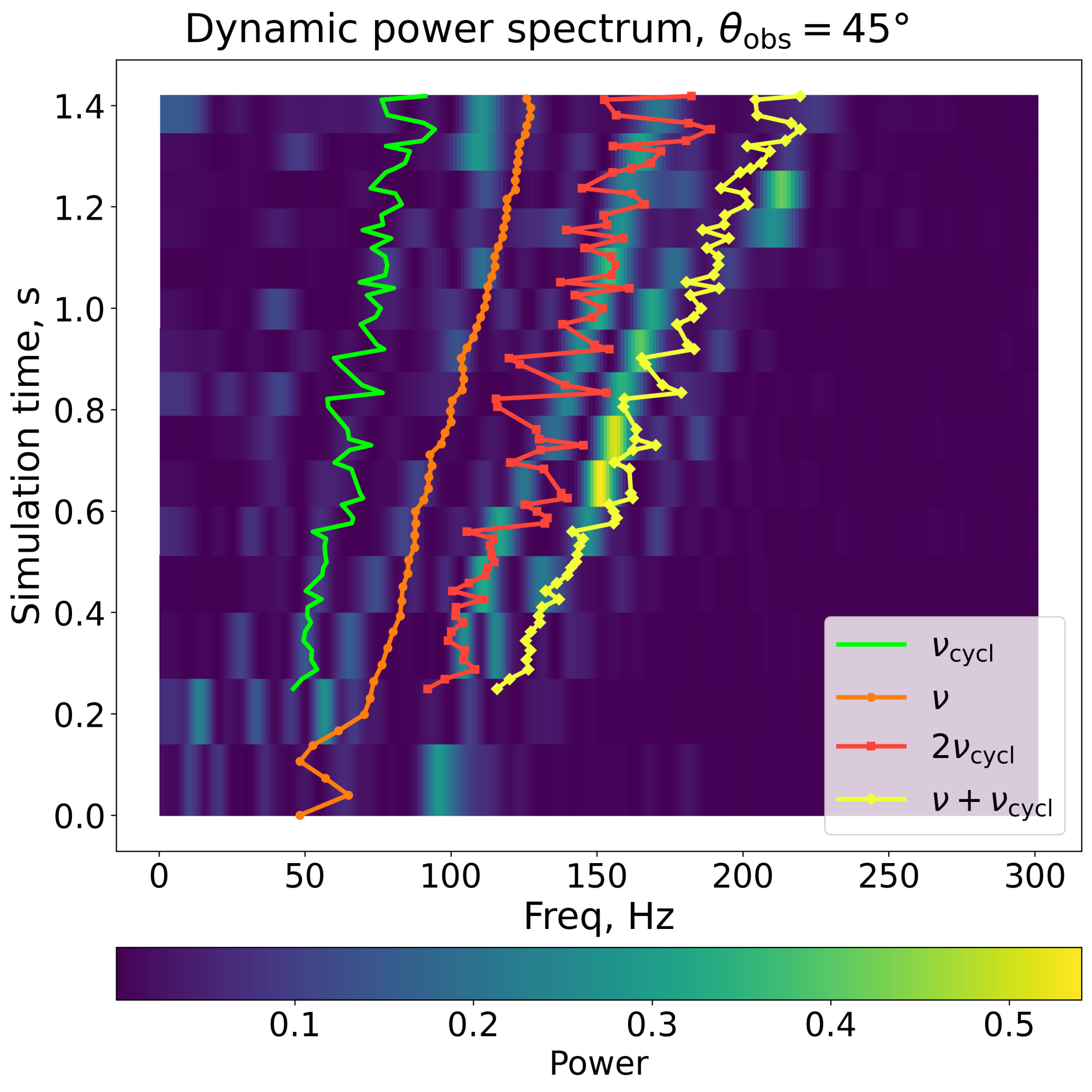}

\vspace{3mm}

\includegraphics[width=0.49\textwidth]{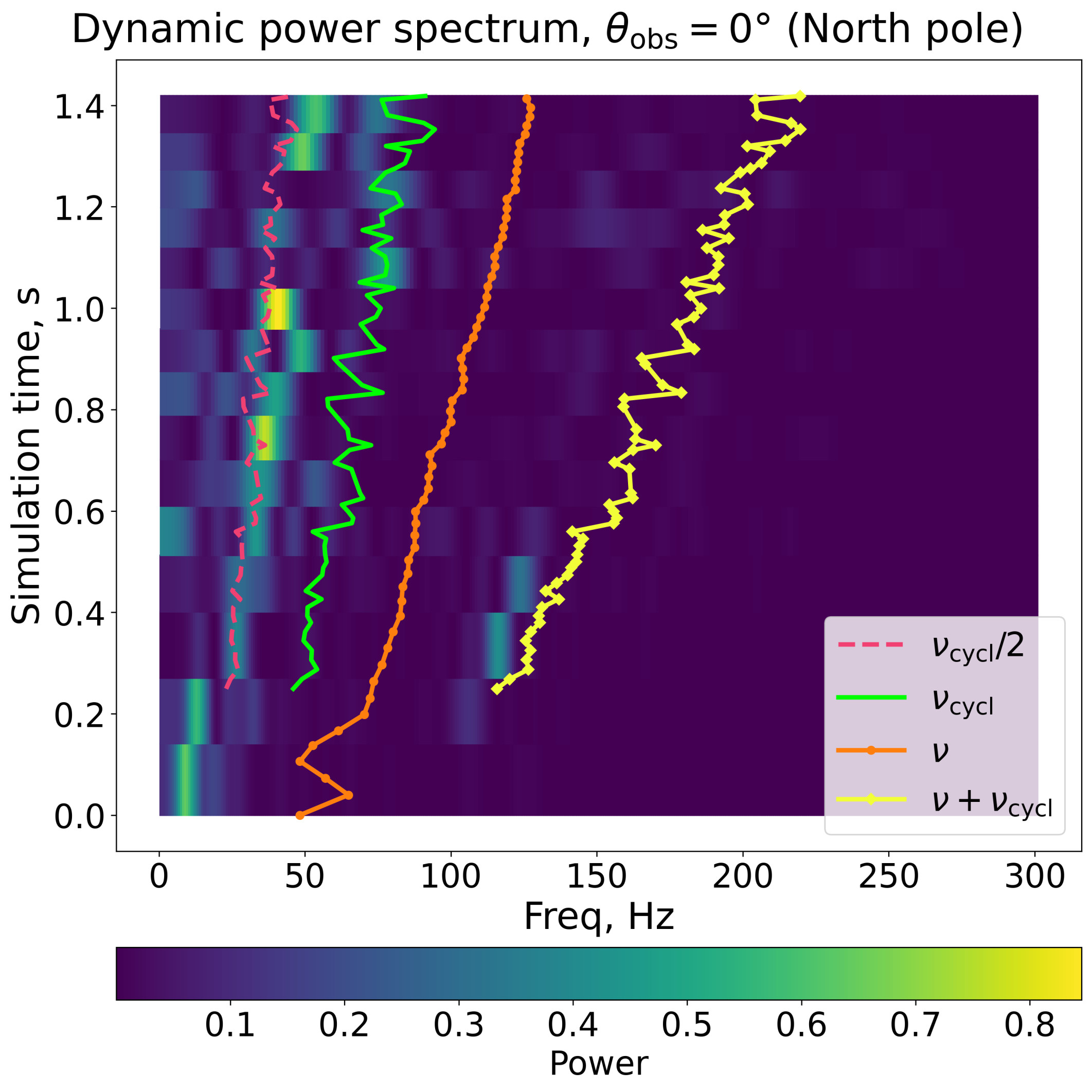}
\includegraphics[width=0.49\textwidth]{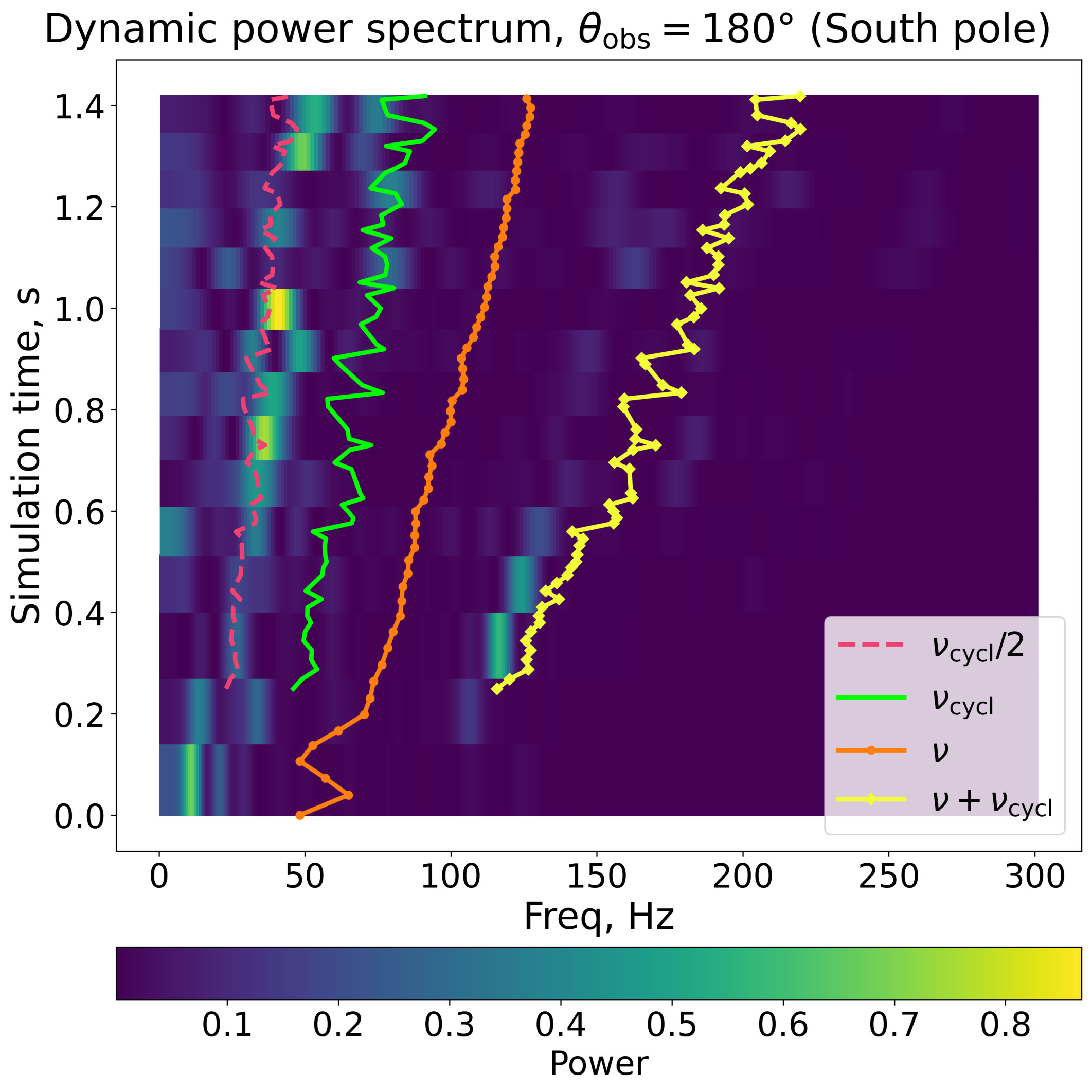}
\end{center}
\caption{Dynamic power spectra of the accretion simulation for different observers. Overplotted are the $z$-component of the average frequency, its first harmonic, cyclone frequency with its first harmonic and the frequency sum.}\label{fig:dyn_pow_sp}
\end{figure*}

The time-resolved Lomb-Scargle periodograms are shown in Figure~\ref{fig:dyn_pow_sp}, where we also plot the mean rotation frequency $\nu$ and the rotation frequency of the cyclones $\nu_{\rm cycl}$ and its lower and higher overtones (see Section~\ref{sec:SL:dyn}).

For the 90\textdegree \ observer, one can see three main variability modes. The first frequency is located at double the average pattern rotation frequency. This signal starts around $t\sim 0.3$s, when the cyclones are formed.
The cyclones are roughly equidistant in longitude and comparable in size and pressure variation, that makes the second harmonic of the rotation frequency dominant. The second mode is visible after the moment $t=1.2$s at a frequency of about 100 Hz, and it can not be attributed neither to the mean rotation frequency nor to the rotation frequency of the pattern. A possible explanation is related to the fact that after a certain point in the simulation, rotation starts to differ between the southern and northern hemispheres (see Figure~\ref{fig:split}). The southern hemisphere, where the cyclones are located, rotates slower than the northern one, and the observed frequency is close to the average rotation frequency of the southern hemisphere and is potentially related to entropic variations in the flow caused by the cyclones. There is also a visible signal between $2\nu_{\rm cycl}$ and $\nu+\nu_{\rm cycl}$ that is more prominent for the observer inclined by 45\textdegree.

The dynamic power density spectrum seen by the observer inclined by 45\textdegree is qualitatively similar to that for the equatorial observer. For the inclinations of 0 and 180\textdegree, the set of characteristic frequencies is different. Most importantly, there are two  low-frequency signals at about  $ \nu_{\rm cycl}/2$ and $ \nu_{\rm cycl}$ dominating the spectrum after $t\sim 0.6$s. These oscillations likely reflect the latitudinal motion of the cyclones, probably being a result of some inertial mode.

% Plotting $\nu_{\rm cycl}$ and $2 \nu_{\rm cycl}$ over the power spectra reveals that the cyclones are indeed the source of low-frequency oscillations for equatorial observer, as seen in Figure \ref{fig:dyn_pow_sp}.

% The polar observer cannot see these; however, he can observe Rossby waves with a frequency of $1.5 \nu$.

%Overall, if we are to look at the plots of conservative variables at different time steps, we can see that after the initial instabilities, the clear sine-wave like pattern forms for both pressure and density. This sine-wave has 2 peaks and it moves faster than the background. It can clearly be seen in the Fig \ref{fig:stream}. 

%\begin{figure*}[h]
%\begin{center}
%\includegraphics[width=0.24\textwidth]{dymamic_power_spec_0c.jpg}
%\includegraphics[width=0.24\textwidth]{dymamic_power_spec_45c.jpg}
%\includegraphics[width=0.24\textwidth]{dymamic_power_spec_90c.jpg}
%\includegraphics[width=0.24\textwidth]{dymamic_power_spec_270c.jpg}
%\end{center}
%\caption{Dynamic power spectra of the accretion simulation for different observers. Cyclon frequency and its first harmonic are overplotted.}\label{fig:dyn_pow_sp_cycl}
%\end{figure*}

%$dE/dt = \frac{d \Sigma}{dt}  (\left( \frac{E}{\Sigma}\right)_d+\frac{1}{2}(\mathbf{v}_{orb}-\mathbf{v})^2)$

\section{Discussion}\label{sec:disc}

% \subsection{Oscillation modes and instabilities}

The results of Section~\ref{sec:SL} show that formation of a differentially rotating spreading layer on the surface of a NS has a set of signature frequencies, dependent on the viewing angle, evolving with time and apparently related to the rotation profile. 
Since rotation is supersonic, one may expect the variability modes to be dominated by Rossby modes. 
Their frequencies are for given latitude equal to $\nu_{\rm R}(\theta) = \nu_{\rm e}(\theta) + m \nu(\theta)$, where $m$ is a whole number,
\begin{equation}
    \nu_{\rm e} = \sqrt{2\nu \ppardir{\theta}{\nu \sin^2\theta}},
\end{equation}
and $\nu = \nu(\theta)$ is the local rotation frequency (see discussion section in \citealt{2020A&A...638A.142A}). 
The observed power density spectrum depends on the contribution of different latitudes to the observed variability. 
One possibility is that the oscillations are excited at the latitude or latitudes where the rotation frequency is equal to the rotation rate of the double-wave pattern. 
For most of the time span, however, the pattern rotation frequency is outside the rotation frequency range within the flow. 
More importantly, Rossby waves may be excited at the location of the cyclones in the Southern hemisphere, and the strongest peaks seen in the lower panels of Figure~\ref{fig:dyn_pow_sp} correspond to the local $\nu_{\rm cycl}$. 
The lower-frequency peaks in the polar periodograms are likely of different origin, as their frequencies do not depend on the rotation frequency during the most of the simulation time span. 

Gravity modes are unlikely to be present in the simulated variability curves, as the configuration starts with neutral convective stability, and the heat released near the equator causes latitudinal mixing that restores an approximately flat entropy profile. 
Pressure modes might be present but confined to latitudinal strips $\sim 1/\Mach$ wide. 
Their frequencies may be estimated as 
\begin{equation}
    \nu_{\rm p} \simeq \frac{c_{\rm s}}{2\uppi R} \simeq \frac{\nu}{\Mach}.
\end{equation}
In our setup, the observed temporal evolution is an outcome of the large velocity difference between the initial rigidly rotating configuration and the newly accreting matter from the disk. 
Depending on the mass accretion rate and cooling rate, this likely causes either Kelvin-Helmholtz (if accretion is fast enough to create a discontinuity in the velocity profile) or Rayleigh-Taylor instability. 
In the simulations we present in Section~\ref{sec:SL}, heating in the equatorial region is much more important. As a result, the excess of entropy near the equator creates favorable conditions for Rayleigh-Taylor instability with the wave numbers $k \lesssim 1 / R \sigma_{\alpha}$. 
For a wave number larger than this value, the instability analysis is done in Appendix~\ref{App:B}.
However, for the long-wavelength modes that dominate the evolution at $t \gtrsim 0.05$s, the instability is classical Rayleigh-Taylor with the effective gravity equal to 
\begin{equation}
    g_{\theta, \, \rm eff} \simeq \Omega^2 R \sin \theta \cos \theta,
\end{equation}
and the growth increment
\begin{equation}
    \sigma_{\rm RT} \simeq \sqrt{g_{\theta, \, \rm eff}k} \simeq \Omega \sqrt{kR \sin \theta \cos \theta}.
\end{equation}
The value varies from $\sim 700$Hz for $k = 1 / R \sigma_{\alpha}$ to $\sim 200$Hz if $k \sim 1/R$, that ensures a rapid growth stage during the first several rotation periods in the SL simulation that stops after entering non-linear regime at $t\sim 0.2$s. 
After that, the density, pressure, and vorticity maps show remarkable stability and rigid-body-like rotation. 
Apparently, the longest instability wave number allowed by the topology of the sphere corresponds to the azimuthal number of $m=2$, that determines the azimuthal structure of the pattern. 
The rotation velocity of the pattern is close to the rotation velocity in the initial conditions, and much smaller than the average velocity of the flow at the same time. 

Starting from $t\sim 0.07$s, there is a measurable difference in the rotation velocities and stream patterns between the Northern and the Southern hemispheres, reaching about 20\% by the end of the simulation (see Figure~\ref{fig:split}). 
One of the most striking differences between the two hemispheres (probably related to the angular frequency asymmetry) is the formation of a pair of large stable cyclones in the Southern hemisphere. 
We suggest that this symmetry breaking is spontaneous and related to the rapid growth of the instability at earlier times. 
It may also be thought as a manifestation of a more global heating instability discussed in \citet{2020A&A...638A.142A}: larger velocity difference in one of the hemispheres leads to larger energy dissipation, the latter creates latitudinal pressure asymmetry that pushes the accreting matter towards the hemisphere already rotating faster. 
Asymmetry in accreted mass would ultimately stop this instability, but the relevant time scale is difficult to estimate because of the large number of factors such as cooling rate, mass sources and sinks, and velocity profiles,  are at play at the same time.
More simulations are needed to check this hypothesis. 

A valid question is if the initial conditions we use correspond to some real stage in the evolution of an accreting NS. 
We start with a gaseous atmosphere rotating as a rigid body at a rate several times less than the Keplerian rate. 
This is an expected outcome of a low-accretion-rate stage when the flow efficiently exchanges angular momentum in latitudinal direction (otherwise, rotation is going to be differential) and slows down due to some kind of friction with the NS crust, presumably rotating even slower. 
Hence, our setup may correspond to an episode of rapidly growing mass accretion rate. 
In future simulations, we are planning to include a friction force and evolve the simulation toward a steady state. 

\section{Conclusions}\label{sec:conc}

The code presented in this work allows to address any possible two-dimensional hydrodynamical problems on a spherical surface up to reasonably high Mach numbers of $\sim 5-10$. 
Both steady-state and dynamical problems set on a sphere show good consistency with the analytical solutions as well as internal consistency between different resolutions and grid types. 

The capabilities of the code allow us for the first time to simulate the dynamics of an accretion spreading layer in the realistic parameter range appropriate for a real weakly magnetized accreting NS. 
Applying the code to the spreading layer problem reveals some interesting properties of this flow. 
In particular, its early evolution is dominated by heating in the equatorial regions, that leads to Rayleigh-Taylor instability and efficient mixing of the accreting matter in latitudinal direction. 
As a by-product, it produces a global two-armed `tennis-ball' pattern which causes flux variations for an observer at a large inclination. 
The pattern retains its shape and changes its rotation velocity at a much slower rate than the spin-up of the spreading layer in general. 
Other observed oscillation modes may be connected to pressure modes propagating in longitudinal direction or to inertial Rossby modes, presumably with the azimuthal numbers $m=0$ and $2$.

Somehow, the development of the instabilities and rotating patterns leads to a considerable asymmetry between the two hemispheres. 
In terms of the average rotation velocities, the asymmetry reaches about 20\% by the end of the simulation run, and is likely to grow further. 
To answer the question if this asymmetry is likely to survive in the steady-state regime, we are planning more simulations with a similar setup but longer duration. 
This will also allow to test the validity of steady-state spreading-layer models.

\begin{acknowledgments}
    This work was supported by a grant from the Simons Foundation (00001470, PA) and NSF-BSF grant 202074. The authors would like to thank Joonas N\"attil\"a and Sergey Khaibrakhmanov for valuable discussions. 
\end{acknowledgments}

\appendix
\section{Analytic solutions for rigid-body rotation}\label{app:A}

Hydrostatic balance in vector form on the surface of a sphere, assuming rigid-body rotation with angular frequency $\Omega$, after vertical integration and projection onto the polar angle direction has the form
\begin{equation}
    \vector{e}_\theta \nabla \Pi = \frac{v^2}{R\sin\theta} \Sigma (\vector{e}_\varpi \cdot \vector{e}_\theta),
\end{equation}
where $\theta$ is polar angle, $\vector{e}_\varpi$ is cylindrical radial unit vector (directed away from the axis of rotation), and $v=\Omega R \sin\theta$ is rotation velocity. 
Projecting the equation onto the polar angle unit vector $\vector{e}_\theta$ yields
\begin{equation}\label{E:App:hydrostatics}
    \partial_\theta \Pi =  \Omega^2 R^2 \Sigma \sin \theta \cos \theta.
\end{equation}
As there is no other equation to constrain the pressure and density profile, it is possible to set an arbitrary profile of density, pressure, or entropy. 
In particular, for $\Sigma = $const, pressure is found as
\begin{equation}
    \Pi_{\rm \Sigma = \const} = \Pi_{\rm poles} + \frac{\Omega^2 R^2\Sigma}{2} \sin^2\theta.
\end{equation}
% \section{Adiabatic constant-entropy solution}\label{app:C}
If globally $\Pi = \Pi_0 \left( \Sigma/\Sigma_0\right)^\Gamma$, solving equation~(\ref{E:App:hydrostatics}) results in
\begin{equation}
   \displaystyle \Sigma_{\rm isentropic} = \Sigma_0 \left[ 1+\frac{\Gamma-1}{2} \Mach_{0}^2 \sin^2\theta\right]^{\frac{1}{\Gamma-1}},
\end{equation}
\begin{equation}
   \displaystyle \Pi_{\rm isentropic} = \Pi_0 \left[ 1+\frac{\Gamma-1}{2} \Mach_{0}^2 \sin^2\theta\right]^{\frac{\Gamma}{\Gamma-1}},
\end{equation}
where the index `0` corresponds to the poles, and $\Mach_0 = \frac{\Omega R}{a_0}$, $a_0 = \sqrt{\frac{\Gamma \Pi_0}{\Sigma_0}}$.
Pressure normalization may be expressed using other quantities as
\begin{equation}
    \Pi_0 = \frac{\Sigma_0}{\Gamma \Mach_0^2} \Omega^2 R^2.
\end{equation}
The Mach number of the isentropic flow is
\begin{equation}
    \Mach(\theta) = \Mach_0 \frac{\sin \theta}{\sqrt{1+\frac{\Gamma-1}{2} \Mach_{0}^2 \sin^2\theta}}.
\end{equation}
It is maximal at the equator, where 
\begin{equation}
    \Mach_{\rm eq} = \frac{\Mach_0}{\sqrt{1+\frac{\Gamma-1}{2} \Mach_{0}^2}}.
\end{equation}
The upper limit for the Mach number for a rigidly rotating isentropic solution is $\Mach_{\rm eq, \ max} = \sqrt{2/(\Gamma-1)} = \sqrt{8} \simeq 2.8$. 

Depending on the latitudinal entropy gradient, a rigidly rotating configuration may be stable or unstable to buoyancy modes \citep{1961hhs..book.....C}.
The derivation in Boussinesq approximation is given in App.~\ref{App:B} and implies that the constant-density case should show convective instability with the increment $\sim \Mach \Omega /2$. 

\section{Stability of rigid-body rotation}\label{App:B}

Let us consider steady-state rigid-body rotation flow from App.~\ref{app:A} and perform its linear analysis in Boussinesq approximation \citep[chapter II]{1961hhs..book.....C}. 
The equation of state is considered adiabatic with the index of $\Gamma$, hence we can introduce entropy as
\begin{equation}
    S = \ln \frac{\Pi}{\Sigma^\Gamma}.
\end{equation}
The unperturbed solution is described by equation~(\ref{E:App:hydrostatics}). 
In the first order, we will need to linearize the continuity equation
\begin{equation}\label{E:general:continuity}
    \partial_t \Sigma + \nabla (\Sigma \vector{v}) = 0,
\end{equation}
Euler equation (2 tangential components)
\begin{equation}\label{E:general:Euler}
    \partial_t \vector{v} + (\vector{v} \nabla) \vector{v} = \vector{g} - \frac{1}{\Sigma} \nabla \Pi,
\end{equation}
and equation of state
\begin{equation}\label{E:general:entropy}
    \partial_t S + (\vector{v} \nabla) S = 0.
\end{equation}
The background solution is determined only by the $\theta$ component of the Euler equation, as the flow is stationary and incompressible, and equations~(\ref{E:general:continuity}) and (\ref{E:general:entropy}) are fulfilled independently of the pressure and density profile.

Let us consider linear perturbations of $\Pi$, $\Sigma$, and $v_{\theta, \ \varphi}$ in the co-rotating frame. For every quantity $f$, $ \delta f \propto \exp\left(\i \left( \omega t - m \varphi \right) \right) f_\theta(\theta)$. 
As we consider the problem in the co-rotating frame, both velocity components are zero in zeroth order, and $v_\theta$ and $v_\varphi$ will be used without deltas. 
Thermodynamical quantities $\Pi$ and $\Sigma$ without deltas in this section will refer to the background solution. 

Linearized continuity equation
\begin{equation}\label{E:lin:continuity}
 \displaystyle  \i \omega R\sin \theta \frac{\delta \Sigma}{\Sigma} - \i m v_\varphi + \partial_\theta \left( v_\theta \sin\theta \right) = 0.
\end{equation}
Linearized entropy conservation equation
\begin{equation}\label{E:lin:entropy}
    \i \omega R \left( \frac{\delta \Pi }{\Pi } - \Gamma \frac{\delta \Sigma}{\Sigma}\right) + v_\theta \partial_\theta S = 0.
\end{equation}
Linearized $\theta$ component of Euler
\begin{equation}\label{E:lin:Euler:th}
    \i \omega R v_\theta = \frac{\delta \Sigma}{\Sigma} \Omega^2R^2 \sin \theta \cos\theta - \frac{1}{\Sigma} \partial_\theta (\delta \Pi). 
\end{equation}

If we ignore pressure variations (that corresponds to the Boussinesq approximation, valid for the processes slow enough for pressure to be leveled by sonic waves), equations~(\ref{E:lin:entropy}) and (\ref{E:lin:Euler:th}) may be combined to reproduce the local Brunt-V\"ais\"al\"a frequency $N$ as
\begin{equation}
    \frac{N^2}{\Omega^2} = - \partial_\theta S \sin\theta \cos \theta.
\end{equation}
The right-hand side of the equation is always negative if entropy increases towards the equator. 
In particular, for $\Sigma = $const, 
\begin{equation}
    \partial_\theta S = \frac{\partial_\theta \Pi}{\Pi} = \frac{\Omega^2 R^2 \Sigma}{\Pi} \sin \theta \cos \theta =  \frac{\Gamma \Mach_0^^2 \sin \theta \cos \theta}{1 + \Gamma \Mach_0^2 \sin^2/2}
    % \Gamma \Mach^2 \sin \theta \cos \theta \left[ 1 + \frac{\Gamma-1}{2} \Mach^2 \sin^2\theta\right]^{-\frac{\Gamma}{\Gamma-1}},
\end{equation}
where $\Mach_0 = \sqrt{\Gamma \Pi_{\rm poles} / \Sigma}$, hence
\begin{equation}
    N = \i \Omega \sqrt{\Gamma} \Mach_0 \frac{\left|\sin \theta \cos \theta \right|}{ \sqrt{ 1 + \Gamma \Mach_0^2 \sin^2\theta/2}}.
\end{equation}
The frequency is always imaginary, meaning convective stability on all the latitudes. 
The increment itself is, however, a strong function of latitude.
For $\Mach \ll 1$, the maximal value is achieved at $\theta \simeq \pi/4$ and equals ${\rm Im}\, \omega \simeq \sqrt{\Gamma} \Mach \Omega / 2$. 

\bibliography{sample631}{}

\hyphenation{Post-Script Sprin-ger}
\begin{thebibliography}{}
\expandafter\ifx\csname natexlab\endcsname\relax\def\natexlab#1{#1}\fi
\providecommand{\url}[1]{\href{#1}{#1}}
\providecommand{\dodoi}[1]{doi:~\href{http://doi.org/#1}{\nolinkurl{#1}}}
\providecommand{\doeprint}[1]{\href{http://ascl.net/#1}{\nolinkurl{http://ascl.net/#1}}}
\providecommand{\doarXiv}[1]{\href{https://arxiv.org/abs/#1}{\nolinkurl{https://arxiv.org/abs/#1}}}

\bibitem[{{Abolmasov} {et~al.}(2020){Abolmasov}, {N{\"a}ttil{\"a}}, \& {Poutanen}}]{2020A&A...638A.142A}
{Abolmasov}, P., {N{\"a}ttil{\"a}}, J., \& {Poutanen}, J. 2020, \aap, 638, A142, \dodoi{10.1051/0004-6361/201936958}

\bibitem[{{Abolmasov} \& {Poutanen}(2021)}]{2021A&A...647A..45A}
{Abolmasov}, P., \& {Poutanen}, J. 2021, \aap, 647, A45, \dodoi{10.1051/0004-6361/202039485}

\bibitem[{{Bahramian} \& {Degenaar}(2023)}]{2023hxga.book..120B}
{Bahramian}, A., \& {Degenaar}, N. 2023, in Handbook of X-ray and Gamma-ray Astrophysics, 120, \dodoi{10.1007/978-981-16-4544-0_94-1}

\bibitem[{{Bransgrove} {et~al.}(2018){Bransgrove}, {Levin}, \& {Beloborodov}}]{2018MNRAS.473.2771B}
{Bransgrove}, A., {Levin}, Y., \& {Beloborodov}, A. 2018, \mnras, 473, 2771, \dodoi{10.1093/mnras/stx2508}

\bibitem[{{Chandrasekhar}(1961)}]{1961hhs..book.....C}
{Chandrasekhar}, S. 1961, {Hydrodynamic and hydromagnetic stability}

\bibitem[{Chen {et~al.}(2020)Chen, Lin, Li, \& Yan}]{doi:10.1137/18M119032X}
Chen, S., Lin, B., Li, Y., \& Yan, C. 2020, SIAM Journal on Scientific Computing, 42, B921, \dodoi{10.1137/18M119032X}

\bibitem[{Gall(1885)}]{gall1885use}
Gall, J. 1885, Scottish Geographical Magazine, 1, 119

\bibitem[{{Gilfanov} {et~al.}(2003){Gilfanov}, {Revnivtsev}, \& {Molkov}}]{2003A&A...410..217G}
{Gilfanov}, M., {Revnivtsev}, M., \& {Molkov}, S. 2003, \aap, 410, 217, \dodoi{10.1051/0004-6361:20031141}

\bibitem[{{Gottlieb} \& {Shu}(1997)}]{1997SIAMR..39..644G}
{Gottlieb}, D., \& {Shu}, C.-W. 1997, SIAM Review, 39, 644, \dodoi{10.1137/S0036144596301390}

\bibitem[{{Hasinger} \& {van der Klis}(1989)}]{1989A&A...225...79H}
{Hasinger}, G., \& {van der Klis}, M. 1989, \aap, 225, 79

\bibitem[{{Ingram} \& {Motta}(2019)}]{2019NewAR..8501524I}
{Ingram}, A.~R., \& {Motta}, S.~E. 2019, \nar, 85, 101524, \dodoi{10.1016/j.newar.2020.101524}

\bibitem[{{Inogamov} \& {Sunyaev}(1999)}]{1999AstL...25..269I}
{Inogamov}, N.~A., \& {Sunyaev}, R.~A. 1999, Astronomy Letters, 25, 269, \dodoi{10.48550/arXiv.astro-ph/9904333}

\bibitem[{{Inogamov} \& {Sunyaev}(2010)}]{2010AstL...36..848I}
---. 2010, Astronomy Letters, 36, 848, \dodoi{10.1134/S1063773710120029}

\bibitem[{{Jakob-Chien} {et~al.}(1995){Jakob-Chien}, {Hack}, \& {Williamson}}]{1995JCoPh.119..164J}
{Jakob-Chien}, R., {Hack}, J.~J., \& {Williamson}, D.~L. 1995, Journal of Computational Physics, 119, 164, \dodoi{10.1006/jcph.1995.1125}

\bibitem[{{Kluzniak} {et~al.}(1990){Kluzniak}, {Michelson}, \& {Wagoner}}]{1990ApJ...358..538K}
{Kluzniak}, W., {Michelson}, P., \& {Wagoner}, R.~V. 1990, \apj, 358, 538, \dodoi{10.1086/169006}

\bibitem[{{Landau} \& {Lifshitz}(1987)}]{landafshitz}
{Landau}, L.~D., \& {Lifshitz}, E.~M. 1987, {Fluid Mechanics}

\bibitem[{{Lomb}(1976)}]{1976Ap&SS..39..447L}
{Lomb}, N.~R. 1976, \apss, 39, 447, \dodoi{10.1007/BF00648343}

\bibitem[{{N{\"a}ttil{\"a}} {et~al.}(2024){N{\"a}ttil{\"a}}, {Cho}, {Skinner}, {Most}, \& {Ripperda}}]{2024ApJ...971...37N}
{N{\"a}ttil{\"a}}, J., {Cho}, J. Y.~K., {Skinner}, J.~W., {Most}, E.~R., \& {Ripperda}, B. 2024, \apj, 971, 37, \dodoi{10.3847/1538-4357/ad54c2}

\bibitem[{{Papaloizou} \& {Stanley}(1986)}]{1986MNRAS.220..593P}
{Papaloizou}, J.~C.~B., \& {Stanley}, G.~Q.~G. 1986, \mnras, 220, 593, \dodoi{10.1093/mnras/220.3.593}

\bibitem[{{Payne} \& {Melatos}(2004)}]{2004MNRAS.351..569P}
{Payne}, D.~J.~B., \& {Melatos}, A. 2004, \mnras, 351, 569, \dodoi{10.1111/j.1365-2966.2004.07798.x}

\bibitem[{{Scargle}(1982)}]{1982ApJ...263..835S}
{Scargle}, J.~D. 1982, \apj, 263, 835, \dodoi{10.1086/160554}

\bibitem[{{Shakura} \& {Sunyaev}(1973)}]{1973A&A....24..337S}
{Shakura}, N.~I., \& {Sunyaev}, R.~A. 1973, \aap, 24, 337

\bibitem[{{Shakura} \& {Sunyaev}(1988)}]{1988AdSpR...8b.135S}
---. 1988, Advances in Space Research, 8, 135, \dodoi{10.1016/0273-1177(88)90396-1}

\bibitem[{Shu(2020)}]{Shu_2020}
Shu, C.-W. 2020, Acta Numerica, 29, 701–762, \dodoi{10.1017/S0962492920000057}

\bibitem[{Sieger(2021)}]{sieger_2021_generating}
Sieger, D. 2021, Generating Meshes of a Sphere.
\newblock \url{https://danielsieger.com/blog/2021/03/27/generating-spheres.html}

\bibitem[{{Suleimanov} \& {Poutanen}(2006)}]{2006MNRAS.369.2036S}
{Suleimanov}, V., \& {Poutanen}, J. 2006, \mnras, 369, 2036, \dodoi{10.1111/j.1365-2966.2006.10454.x}

\bibitem[{Touze {et~al.}(2015)Touze, Murrone, \& Guillard}]{cletouze_2015_multislope}
Touze, C.~L., Murrone, A., \& Guillard, H. 2015, Journal of Computational Physics, 284, 389, \dodoi{10.1016/j.jcp.2014.12.032}

\bibitem[{{Ursini} {et~al.}(2024){Ursini}, {Gnarini}, {Capitanio}, {Bobrikova}, {Cocchi}, {Di Marco}, {Fabiani}, {Farinelli}, {La Monaca}, {Rankin}, {Saade}, \& {Poutanen}}]{2024Galax..12...43U}
{Ursini}, F., {Gnarini}, A., {Capitanio}, F., {et~al.} 2024, Galaxies, 12, 43, \dodoi{10.3390/galaxies12040043}

\bibitem[{{van der Klis}(2000)}]{2000ARA&A..38..717V}
{van der Klis}, M. 2000, \araa, 38, 717, \dodoi{10.1146/annurev.astro.38.1.717}

\bibitem[{{van der Klis}(2001)}]{2001ApJ...561..943V}
---. 2001, \apj, 561, 943, \dodoi{10.1086/323378}

\bibitem[{{van Leer}(1979)}]{VANLEER1979101}
{van Leer}, B. 1979, Journal of Computational Physics, 32, 101, \dodoi{https://doi.org/10.1016/0021-9991(79)90145-1}

\bibitem[{{Watts} {et~al.}(2003){Watts}, {Andersson}, {Beyer}, \& {Schutz}}]{2003MNRAS.342.1156W}
{Watts}, A.~L., {Andersson}, N., {Beyer}, H., \& {Schutz}, B.~F. 2003, \mnras, 342, 1156, \dodoi{10.1046/j.1365-8711.2003.06612.x}

\end{thebibliography}
\bibliographystyle{aasjournal}



\end{document}